\documentclass[aps,prb,twocolumn,natbib,showpacs,floatfix,superscriptaddress, longbibliography]{revtex4-2}

\usepackage{graphicx}
\usepackage{amsmath,amsthm,amssymb,amsfonts,bbold,bm,color,mathtools}
\usepackage{float}
\usepackage{epstopdf}
\usepackage{hyperref}
\usepackage{enumerate}
\usepackage{wasysym}
\usepackage{url}
\usepackage{dsfont}
\usepackage{braket}
\usepackage{comment} 
\usepackage[capitalize]{cleveref}
\usepackage[normalem]{ulem}
\usepackage{xcolor}
\usepackage{subfigure}
\usepackage{overpic}
\usepackage{svg}
\graphicspath{{figures/}}
\hypersetup{colorlinks=true,linkcolor=blue,citecolor=blue,urlcolor=blue}

\begin{document}
\newcommand{\pv}[1]{\textcolor{black}{#1}}
\newcommand{\jt}[1]{\textcolor{black}{#1}}
\newcommand{\red}[1]{\textcolor{black}{#1}}
\newcommand{\bscco}{Bi$_2$Sr$_2$CaCu$_2$O$_{8+x}$}

\title{Dynamical signatures and control of time-reversal breaking in twisted nodal superconductors}

\author{Jefferson Tang}
\affiliation{Department of Physics, University of Connecticut, Storrs, Connecticut 06269, USA}
\author{Pavel A. Volkov}
\affiliation{Department of Physics, University of Connecticut, Storrs, Connecticut 06269, USA}
\begin{abstract}
Recent observations of time-reversal breaking superconductivity at twisted cuprate interfaces motivate the development of new approaches to better characterize this emergent phenomenon. Here we study the dynamical properties of the order parameters at the twisted unconventional superconductor interfaces. We reveal the emergence of a soft collective mode (Josephson plasmon) at the time-reversal breaking transition, which can be tuned by temperature, twist angle or magnetic field. Furthermore, nonlinear dynamical responses contain direct signatures of both the transition and the broken symmetry itself.  In particular, we show that \pv{the generation of a second harmonic voltage under alternating current driving} is a necessary and sufficient signature of time-reversal symmetry breaking. Finally, we demonstrate that strong nonlinear driving induces dynamical phase transitions between phases with and without spontaneous symmetry breaking, introducing a tool for their out-of-equilibrium control. We discuss the signatures of our predictions in AC current-driven experiments on twisted \bscco interfaces.
\end{abstract}
\maketitle

\section{Introduction}

Twisted nodal superconductors (see \cite{pixley2025rev} and references therein) have emerged recently as a new platform to realize topological \cite{Can2021,VolkovPRL-2023} and time-reversal breaking (TRSB) \cite{Can2021,Tummuru-Franz-2021,volkov_2025,volkov_diode} superconductivity. For interfaces between d-wave superconductors, a continuous phase transition breaking time-reversal symmetry has been predicted to occur close to 45$^\circ$ twist \cite{Can2021,volkov_2025}. Recent experiments on twisted flakes of \bscco \cite{Zhao2023} demonstrated a trainable Josephson diode effect \cite{volkov_diode}, consistent with spontaneously broken time-reversal symmetry. However, diode effect is sufficient, but not necessary criterion for TRSB \cite{volkov_diode}. Furthermore, the incomplete diode reversal by training in experiments \cite{Zhao2023} suggest the possibility of an additional time-reversal symmetry breaking being present at the interface. Superconducting field-free diode effect has been also reported for  45$^\circ$ twisted \bscco interfaces ~\cite{xue_2023_OP} (which is unexpected theoretically \cite{volkov_diode}) and in single flakes of \bscco~\cite{diode_1flake_2025}. 
Thus, developing new methods of probing time-reversal breaking is an important challenge for the field \cite{pixley2025rev}.




One of the universal tools for probing symmetry-breaking phase transitions is the spectroscopic detection of the associated soft modes. Their existence is intimately connected to the spontaneous nature of the symmetry breaking \cite{scott_1974} and determines the dynamical behavior of the system close to the transition \cite{halperinhohenberg}. 
This approach has been applied to a wide range of systems: ferroelectrics and multiferroics \cite{scott_1974,Kamba2021}, excitonic insulators \cite{TNS,tns_kim,tnsmai,tnsx,Kogar2017} and frustrated magnets \cite{Giamarchi2008}. Moreover, soft modes have been predicted to accompany \pv{TRSB \cite{lin_2012,stanev_2012,babaev_2011,maiti2013,marciani2013}} transitions in bulk superconductors, with some experimental signatures having been reported \cite{trsb_fe}.



Another technique that has recently found application in superconductivity research is nonlinear response. In particular, third harmonic generation has been applied to probe collective modes of superconductors \cite{shimano2020}, \pv{including collective phase modes - plasmons \cite{Gabriele2021,PhysRevB.110.L060504,Katsumi2023,Rajasekaran2018,Salvador2024,Kaj2023,Seibold2021,Zhang2023NSR}}, while second harmonic generation and photocurrent has been proposed to probe the quantum geometric properties of quasiparticles \cite{fiete2024,kaplan2025}, including applications to twisted nodal superconductors \cite{kaplan2025}. However, nonlinear responses of the order parameter in the TRSB phase of twisted nodal superconductors have not been considered so far.


\begin{figure}
    \centering
    \includegraphics[width=0.95\linewidth]{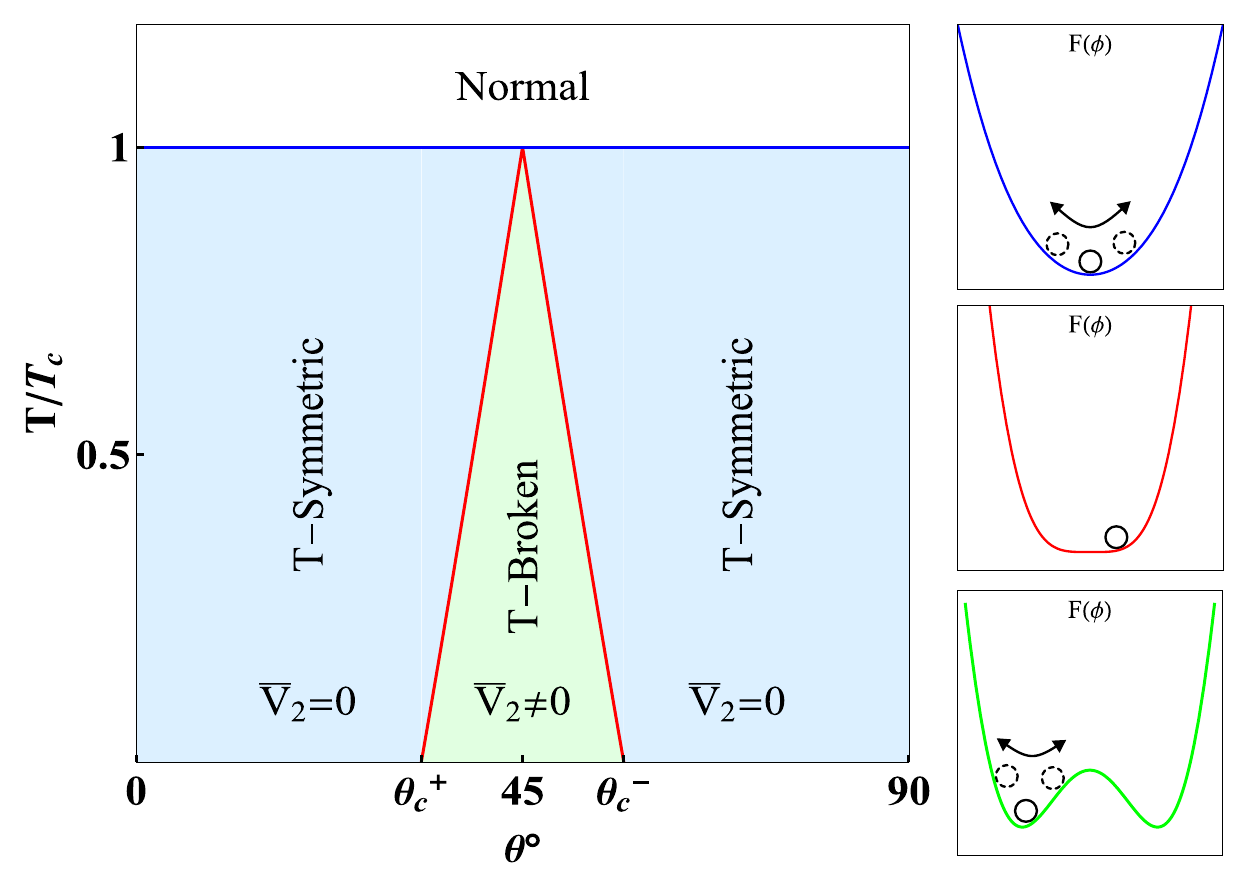}
    \caption{Phase diagram of the interface for varied temperature and twist angle with the panels demonstrating the qualitative behavior of the collective order parameter oscillations (Sec. \ref{Dynamical Behavior}). A soft mode appears at the phase transition point (Sec. \ref{softmode}) and the system generates second harmonic response in the TRSB phase (Sec. \ref{sec: reentrance}).}
    \label{fig:phase-diagram1}
\end{figure}

In this work, we study the collective dynamics of the twisted interfaces between nodal superconductors in vicinity of of the TRSB transition. At the transition (Fig. \ref{fig:phase-diagram1}), we show the emergence of a soft Josephson plasmon with signatures in AC impedance measurements. We further demonstrate that the nonlinear response of this mode within the ordered phase provides an unambiguous signature of the broken symmetry: second-harmonic voltage generation. Increasing the AC drive leads to a sequence of non-equilibrium phase transitions between symmetry breaking and symmetric phases, distinguished by the the presence or absence of the second harmonic. Finally, we provide experimental predictions for twisted interfaces of \bscco.

The paper is organized as follows. In Sec. \ref{model}, we introduce the model for twisted nodal superconductor interfaces, describing both its equilibrium (Sec. \ref{sec:equil}) and dynamical properties (Sec. \ref{Dynamical Behavior}). We investigate collective modes in Sec. \ref{softmode} and their experimental signatures in linear response in \ref{sec:linimped}. In Sec. \ref{sec: reentrance} we study the nonlinear dynamical responses perturbatively, emphasizing the role of second harmonic generation. In Sec. \ref{sec:non-eq} we demonstrate that strong external driving can lead to dynamical phase transitions in the system and characterize the resulting non-equilibrium phase diagram. We discuss the experimental implications of our results and present an outlook in Sec. \ref{sec:outlook}.

\section{Model}\label{model}

\subsection{Equilibrium Properties}

\label{sec:equil}


We first review the equilibrium description of twisted interfaces of nodal superconductors, focusing on the order parameter physics \cite{pixley2025rev}. The two twisted superconducting flakes are described by the respective order parameters $|\Delta_{a,b}| e^{i \red{\phi}_{a,b}}$. While the interactions within each flake fix their amplitudes ($|\Delta_{a}| = |\Delta_{b}|$ for identical superconductors), the phase difference $\red{\phi} = \red{\phi}_a-\red{\phi}_b$ between the two order parameters is determined by the interface. Specifically, the interfacial free energy \cite{Can2021,pixley2025rev} for weak interlayer coupling can be written as:
\begin{equation}\label{free energy}
    F(\phi\pv{,\theta, T}) = F_0 - \frac{\hbar}{2e}\left[J_{c1}\pv{(\theta, T)}\cos(\phi)-\frac{J_{c2}\pv{(\theta,T)}}{2}\cos(2\phi) \right],
\end{equation}
\pv{where the dependence of $J_{c1,2}(\theta, T)$ on $\theta$ and $T$ will be discussed below. We note that this expression can arise from different microscopic mechanisms, such as higher-order tunneling \cite{volkov_2025,Can2021} or disorder \cite{yuan2023,yuan2024}. Our theory, built on Eq. \eqref{free energy}, therefore does not rely on a specific microscopic mechanism and applies to both cases.}
Given the values of $J_{c1}\red{(\theta,T)}$ and $J_{c2}\red{(\theta,T)}$, one can find the equilibrium value of $\phi$ by minimizing Eq. \eqref{free energy} and, as described below, study the system's dynamics. To illustrate qualitatively the dependence of $J_{c1}$ and $J_{c2}$ on physical parameters, we will focus on three: (i) the twist angle $\theta$, (ii) temperature $T$, and (c) in-plane magnetic field. \red{Tuning these parameters result in 2 distinct phases, as we demonstrate below.} 

For the twist angle dependence, we will assume an empirical form $J_{c1}(\theta) = J_{c1} \cos(2 \theta)$; $J_{c2}(\theta) = J_{c2} = const$, which is consistent with d-wave symmetry \cite{pixley2025rev} and recent experiments on \bscco \cite{Zhao2023}. In the vicinity of $\theta =45^\circ$, $J_{c2}$ can be taken as a constant, dominating over $J_{c1}(\theta)$. For $J_{c2}>0$ and $2J_{c2}>|J_{c1}(\theta)|$, \eqref{free energy} has two equivalent minima $\phi_0$ and $-\phi_0$, that break TRS. The critical angle for the onset of the TRSB phase is 
\begin{equation}
    \theta_{c}^{\pm}=\frac{1}{2}\arccos\left(\pm \frac{2J_{c2}}{J_{c1}}\right).
\end{equation}
To describe \pv{qualitatively} the system's temperature dependence, we will adopt the Ginzburg-Landau temperature dependence of the order order parameters $|\Delta_{a,b}(T)| \propto \Delta_0\sqrt{1-T/T_c}$. For weak coupling between flakes across the interface, one finds that \cite{Can2021}: \pv{$J_{c1}(\theta,T) = J_{c1}^{T=0}(\theta)(1-T/T_c)$ and $J_{c2}(T) = J_{c2}^{T=0}(1-T/T_c)^2$.} \red{Taking  $\bar{J}_{c1}^{T=0}=J^{T=0}_{c1}/J^{T=0}_{c2}$,} the critical angle is now given by 
\begin{equation}
\theta_c^{\pm}(T) = \frac{1}{2}\arccos\left(
\pm \frac{2 }{\red{\bar{J}_{c1}^{T=0}}}(1-T/T_c)
\right),
\end{equation}
shown as the red line in Fig. \ref{fig:phase-diagram1}. \pv{Taking the results above into account, in the following we will use for concreteness:
\begin{equation}
\begin{gathered}
    J_{c1}(\theta,T) = J_{c1}^{T=0}\cos(2\theta)(1-T/T_c),
    \\
J_{c2}(T) = J_{c2}^{T=0}(1-T/T_c)^2.
\end{gathered}
\end{equation}
}

Finally, we consider the effect of an in-plane magnetic field $B_\parallel$. Importantly, while it does break TRS, it has different symmetry properties than the order parameter $\phi_0\neq 0$. For two identical flakes an in-plane $C_{2\parallel}$ axis remains a proper symmetry of system even in presence of twist. Under this symmetry, $\phi_0\to-\phi_0$, but $B_\parallel \to B_\parallel$. Thus, in presence of $B_\parallel$, the transition from $\phi=0$ phase to $\phi=\pm\phi_0$ phase remains sharply defined, but the symmetry broken in this case is $C_{2\parallel}$ only \footnote{Note that out-of-plane field does change sign under $C_{2\parallel}$ and breaks the symmetry between $\pm \phi_0$ explicitly \cite{volkov_diode} (see also Sec. \ref{subsec:memory}), that should smear the transition.}. For a sufficiently small interface, $\lambda_J\gg L$ where $\lambda_J$ is the Josephson penetration depth and $L$ is the length of the junction \cite{volkov_2025,barone1982,tinkham2004,volkov_diode} the effect of $B_\parallel$ is as follows. Taking the twisted interface to be in the x-y plane, and taking $B_\parallel$ to be along the $y$ axis, $\phi$ acquires a spatial dependence
\begin{equation}\label{mag phi}
    \phi(x) = \frac{2\pi d}{\Phi_0}B_y x + \phi
\end{equation}
where $d$ is the effective junction thickness, $\Phi_0=hc/(2e)$ the superconducting flux quantum, and $\phi$ is a constant \cite{volkov_2025}. Integrating the free energy, Eq. \eqref{free energy}, with a spatially varying phase (Eq. \eqref{mag phi}) over a square junction with side $L$, we find 
\begin{align}
    F(\phi_0,\bar{\Phi})/L^2 = F_0 &- \frac{\hbar}{2e}\Bigg[J_{c1}(\theta,T)\cos(\phi)\frac{\sin(\pi\bar{\Phi})}{\pi\bar{\Phi}}\nonumber \\
    & -\frac{J_{c2}(T)}{2}\cos(2\phi)\frac{\sin(2\pi\bar{\Phi})}{2\pi\bar{\Phi}}\Bigg].
    \label{eq:Fmag}
\end{align}
where $\bar{\Phi} = dLB_y/\Phi_0$ is the total magnetic flux passing through the junction. For $J\red{^{T=0}}_{c2}>0$, one observes that near $\theta=45^\circ$, where $2J_{c2}\red{(T)}>|J_{c1}|(\theta\red{,T})$, tuning $\bar{\Phi}$ can suppress the $J_{c2}\red{(T)}$ term by a relative factor $\cos(\pi \bar{\Phi})$, allowing one to suppress the symmetry-broken phase and tune through the transition with external field. On the other hand, away from $\theta = 45^\circ$ where $J_{c1}(\theta)> 2J_{c2}$ it is not possible to induce $C_{2\parallel}$ symmetry-breaking starting from the trivial (non-TRSB) phase.

\subsection{Collective dynamics: RCSJ Model}\label{Dynamical Behavior}

We now discuss the description of the low-frequency dynamics of the TRSB order for the twisted superconducting interfaces. The free energy \eqref{free energy} is related to the Josephson current through the interface by the relation \cite{golubov2004}: $I_j(\phi) =\frac{2e}{\hbar}  \partial_{\phi}F$, which reduces in our case to
\begin{equation}
I_j(\phi)
    =
    J_{c1}\red{(\theta,T)}\sin(\phi)-J_{c2}\red{(T)}\sin(2\phi).
    \label{eq:joscur}
\end{equation}
To study the dynamical behavior of the system, we will use the resistively and capacitively shunted junction (RCSJ) model \cite{tinkham2004,barone1982,volkov_diode}
\begin{align}\label{RCSJ model}
    \frac{\hbar C}{2e}\partial_{tt}\phi(t)+\frac{\hbar}{2eR}\partial_t\phi &+ J_{c1}\red{(\theta,T)}\sin(\phi) \nonumber \\
    &-J_{c2}\red{(T)}\sin(2\phi)=I_0(t)
\end{align}
with the voltage $V$ across the interface is given by the Josephson relation $V=\frac{\hbar}{2e}\partial_t\phi$ and $I_0(t)$ is the external driving current. This model is sufficient to describe long-wavelength dynamical properties of the twisted interface both in linear and nonlinear regime \cite{barone1982,tinkham2004}. We further assume that the characteristic frequencies and currents of the twisted interface are much smaller than the bulk ones (consistent with experiments \cite{Zhao2023}), such that the collective modes of the bulk flakes \cite{Savelev2010,Gabriele2021} need not be considered.

In what follows we will specifically focus on the harmonic driving case  $I_0(t) = I_0 \cos(\omega t)$, which allows to probe the response of the symmetry-breaking parameter $\phi$ at a given frequency $\omega$. We can rewrite Eq. \eqref{RCSJ model} in terms of a dimensionless time variable $\tau$:
\begin{align}\label{Normalized RCSJ model}
\partial_{\tau\tau}\phi(\tau) + \frac{1}{\sqrt{\beta_c\red{(T)}}}&\partial_\tau\phi + \bar{J}_{c1}(\theta\red{,T})\sin(\phi) \nonumber \\
&-\red{\left(1-\frac{T}{T_c}\right)^2}\sin(2\phi) =    \bar{I}_{0}(\tau\red{,T}),
\end{align}
where $\beta_c\red{(T)} = 2eR^2CJ_{c2}\red{(T)}/\hbar$, $t_0 = \hbar/(2eRJ_{c2}\red{^{T=0}})$, $\tau= t/(t_0\sqrt{\beta_c})$, \red{$\bar{J}_{c1}(\theta,T)=J_{c1}(\theta,T)/J_{c2}^{T=0} $}, and $\bar{I}=I/J_{c2}\red{^{T=0}} $.




The model, Eq. \eqref{Normalized RCSJ model} describes the dynamics of the phase difference responsible for the TRSB transition. In what follows, we will first investigate linear and nonlinear responses of the model \eqref{Normalized RCSJ model} perturbatively in $I_0$, expanding $\phi(\tau)$ in a Taylor series in $\epsilon = I_0/J_{c2}\red{^{T=0}}$: $\phi(\tau) = \phi_0+\epsilon \phi_1(\tau) + \epsilon^2\phi_2(\tau)+\ldots$. A corresponding expansion is then performed in the equation \ref{Normalized RCSJ model} around the minimum $\phi=\phi_0$. In the TRSB phase $\phi_0 = \arccos\left(\frac{\bar{J}_{c1}(\theta\red{,T})}{2\red{(1-\frac{T}{T_c})^2}}\right)$, from which we get

\begin{align}\label{taylored rcsj eq}
    \bar{I}_0(\tau\red{,T}) = \partial_{\tau\tau}\phi(\tau) &+ \frac{1}{\sqrt{\beta_c}}\partial_\tau\phi + a(\theta\red{,T})(\phi-\phi_0) \nonumber\\&+b(\theta\red{,T})(\phi-\phi_0)^2+c(\theta\red{,T})(\phi-\phi_0)^3
\end{align}
where
\begin{subequations}
    \begin{align}
    a(\theta\red{,T})&=\bar{J}_{c1}(\theta\red{,T})\cos(\phi_0)-2\red{\left(1-\frac{T}{T_c}\right)^2}\cos(2\phi_0) \\
    b(\theta\red{,T})&= 2\red{\left(1-\frac{T}{T_c}\right)^2}\sin(2\phi_0)-\frac{\bar{J}_{c1}(\theta\red{,T})}{2}\sin(\phi_0) \\
    c(\theta\red{,T})&=\frac{4}{3}\red{\left(1-\frac{T}{T_c}\right)^2}\cos(2\phi_0)-\frac{\bar{J}_{c1}(\theta\red{,T})}{6}\cos(\phi_0).
    \end{align}
\end{subequations}
The perturbative expansion for the time-reversal symmetric case is obtained by setting $\phi_0 = 0$. 

To solve for $\phi(\tau)$ perturbatively, we will insert the expansion of $\phi(\tau)$ in $\epsilon$ into \eqref{taylored rcsj eq} and group terms of the same order in $\epsilon$. For example, the equations corresponding to the first two orders in $\epsilon$ are (for harmonic driving $I_0(\tau) = I_0 \cos(\bar{\omega} \tau)$, where $\bar{\omega} = \omega t_0 \sqrt{\beta_c}$): 
\begin{align}
    \label{linear response} &\frac{\epsilon}{2}\left(\red{\cos(\bar{\omega}\tau)}\right) = \epsilon\partial_{\tau\tau}\phi_1(\tau)+\frac{\epsilon}{\sqrt{\beta_c}}\partial_\tau\phi_1(\tau)+\epsilon a(\theta\red{,T})\phi_1(\tau) \\
    \label{2nd harm response}&0 = \epsilon^2\partial_{\tau\tau}\phi_2(\tau)+\frac{\epsilon^2}{\sqrt{\beta_c}}\partial_\tau\phi_2(\tau)+\epsilon^2 a(\theta\red{,T})\phi_2(\tau)+\epsilon^2 b(\theta\red{,T})\phi_1^2(\tau).
\end{align}
The resulting equations can then be solved order by order in $\epsilon$.

\section{Critical mode and linear response}
\label{sec:critlin}

In this section we study the linear response of $\phi$ to an AC drive $I_0(\tau)$. We first analyze the $\beta_c\to \infty$ case, allowing to identify the soft mode of the TRSB transition as the Josephson plasmon of the twisted interface. We demonstrate how to detect and measure the effects of additional, explicit TRSB (suggested by the ``memory effect" in experiments \cite{Zhao2023,volkov_diode}) via the soft mode in Sec. \ref{subsec:memory}. In Sec. \ref{sec:linimped}, we show the signatures of the soft mode in linear impedance measurements at finite $\beta_c$.

\subsection{Soft Josephson plasmons}\label{softmode}

\begin{figure}[h]
    \centering
    \subfigure{
    \includegraphics[width=0.95\linewidth]{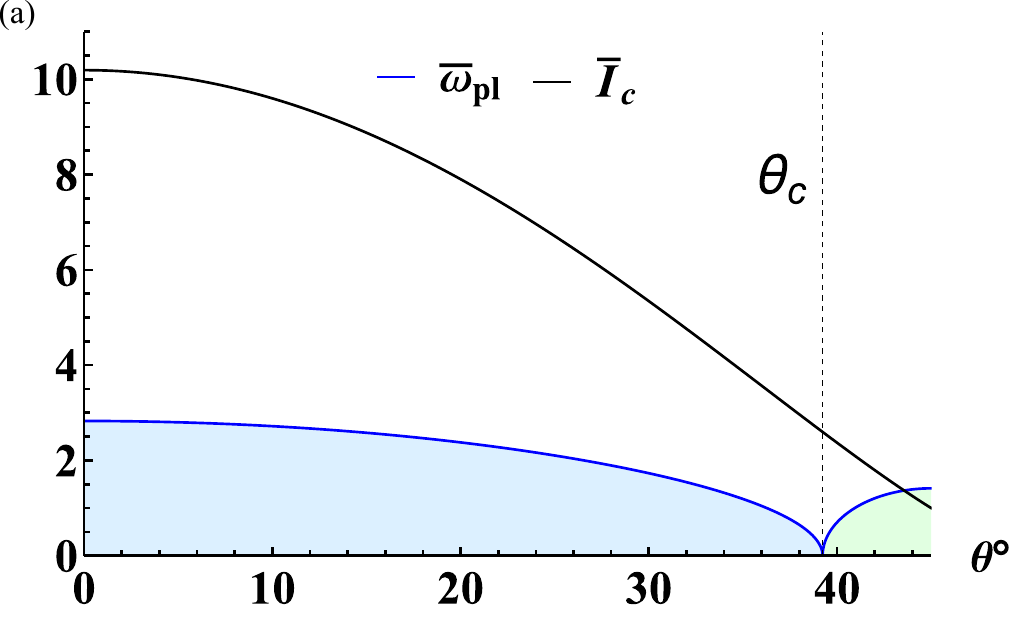}
    \label{fig: plasma vs twist}
    }
    \subfigure{
    \includegraphics[width=0.95\linewidth]{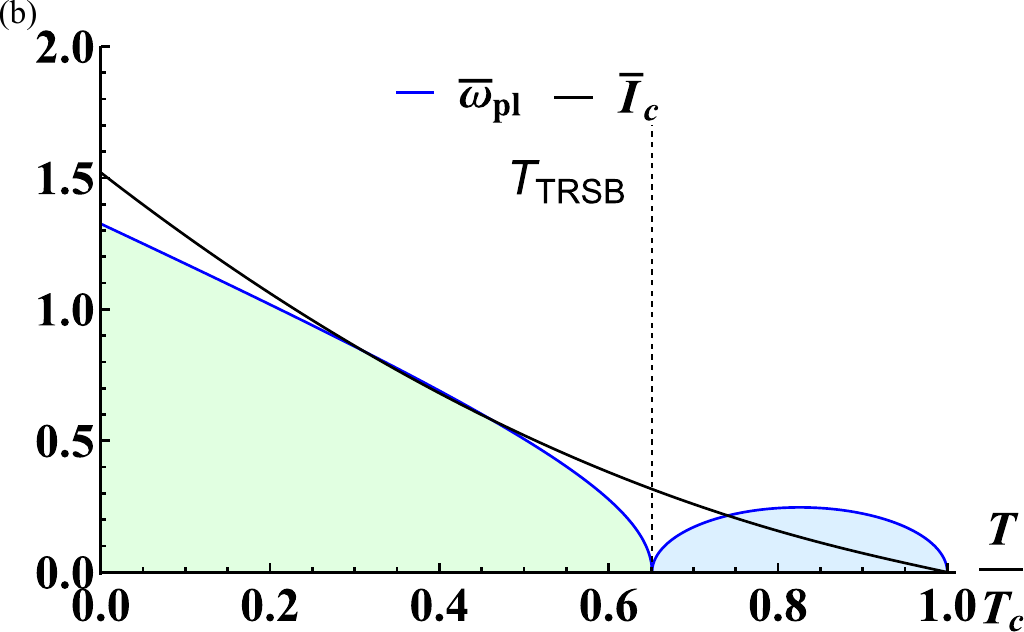}
    \label{fig: plasma vs T}
    }
    \caption{Normalized Josephson plasma frequency $\bar{\omega}_{pl}$, Eq. \eqref{eq:omegapl}, and critical current $\bar{I_c}$ as a function of: a) twist angle $\theta$ and b) temperature (normalized to the critical one), obtained from linearized Eq. \eqref{Normalized RCSJ model} in the $\beta_c\to \infty$ limit. We take $\bar{J}_{c1}^{\red{{T=0}}}=10$, giving $\theta_{c\red{,T=0}}^{+}=39.2^{\circ}$. For, a) $T=0$, and for b) \red{$\theta =43^\circ$ and $T_{TRSB}=0.65T_c$ where $T_{TRSB}$ indicate the onset of time reversal breaking}. Blue filling indicate symmetric $\phi_0 = 0$ phase whereas green corresponds to the TRSB phase.
	}
    \label{fig:plasma vs t,twist}
\end{figure}

We first consider the $\beta_c \to \infty$ limit. Linearizing Eq. \eqref{Normalized RCSJ model} we get an oscillator equation $  \bar{I}_{0}(\tau)=\partial_{\tau\tau}\phi(\tau) +\bar{\omega}_{pl}^2 \phi(\tau)$, where 
\begin{equation}
    \bar{\omega}_{pl}\red{(\theta,T)} = \sqrt{\bar{J}_{c1}(\theta,T)\cos(\phi_0)-2\red{\left(1-\frac{T}{T_c}\right)^2}\cos(2\phi_0)}.
    \label{eq:omegapl}
\end{equation}
This collective excitation of $\phi$, usually called the Josephson plasmon \cite{Savelev2010} has been observed in bulk layered \bscco \cite{plasmon_exp_1999}. In Fig. \ref{fig:plasma vs t,twist} we present the dependence of $\bar{\omega}_{pl}$ on twist or temperature, using the dependencies introduced in Sec. \ref{sec:equil}. 
One notices that $\bar{\omega}_{pl}$ vanishes precisely at the TRSB transition in both cases, as illustrated in Fig. \ref{fig:phase-diagram1}. To understand this behavior, consider the analogy to a harmonic oscillator. The phase transition point corresponds to a flattening of the potential, which can be observed in Fig. \ref{fig:phase-diagram1}, as a result, the harmonic oscillation frequency of $\phi$ vanishes, which suggests an increased role of nonlinear effects, discussed below in Sec. \ref{sec: reentrance}. 

This behavior is in stark contrast to the conventional Josephson junctions. There, the Josephson plasma frequency is proportional to $\sqrt{I_c}$ \cite{shmidt}, $I_c$ being the Josephson critical current $\max_\phi[I_j(\phi)]$ \cite{tinkham2004}. In our case, the critical current is given by \cite{volkov_diode}:  
\begin{equation}
\bar{I}_c =\bar{J}_{c1}(\theta\red{,T})\sin(\phi')-\red{\left(1-\frac{T}{T_c}\right)^2}\sin(2\phi').
\end{equation}
where
\begin{equation}
    \phi' = \arccos\left(\frac{\bar{J}_{c1}(\theta\red{,T})-\sqrt{\bar{J}_{c1}(\theta\red{,T})^2+32\red{\left(1-\frac{T}{T_c}\right)^4}}}{8\red{\left(1-\frac{T}{T_c}\right)^2}}\right).
\end{equation}
The resulting $\bar{I}_c(T,\theta)$, shown in Fig. \ref{fig:plasma vs t,twist}, is monotonic and does not vanish together with $\bar{\omega}_{pl}$ at the TRSB transition, in clear violation of the conventional Josephson junction behavior.

\begin{figure}
    \centering
    \subfigure{\includegraphics[width=0.95\linewidth]{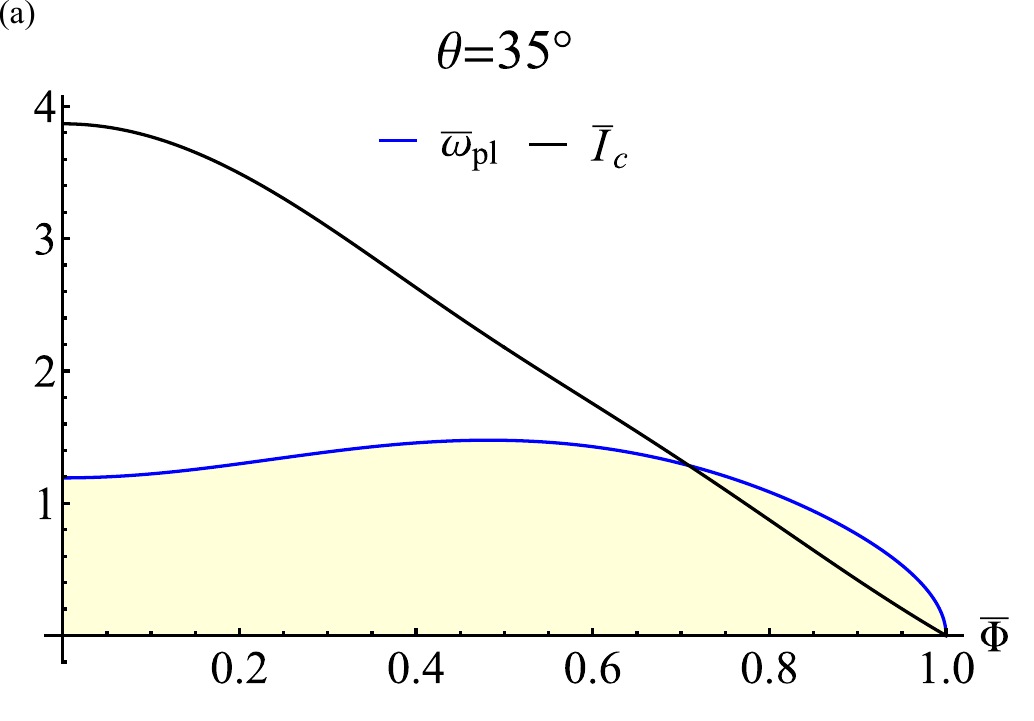}}
    \subfigure{\includegraphics[width=0.95\linewidth]{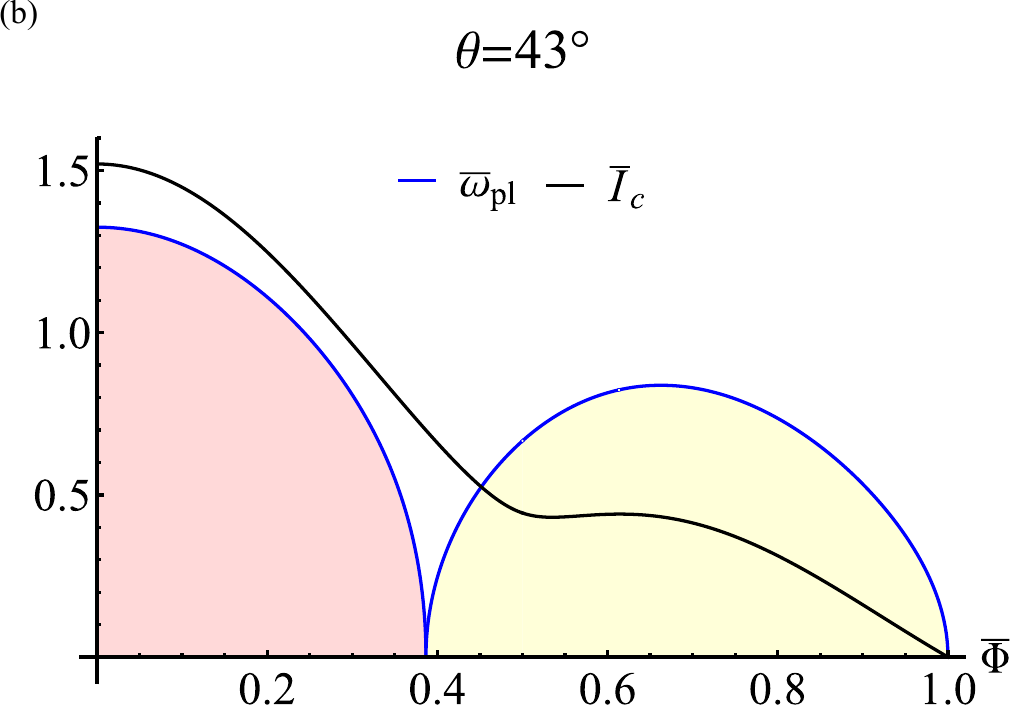}}
    \caption{Normalized Josephson plasma frequency $\bar{\omega}_{pl}$, Eq. \eqref{eq:omegapl}, and critical current $\bar{I_c}$ as a function of in-plane magnetic flux $\bar{\Phi}$ through the system (see Eq. \eqref{eq:Fmag})
    for $\bar{J}_{c1}^{\red{{T=0}}}=10$ \red{and $T=0$}. The yellow filling indicates that the system exhibit $C_{2y}$ symmetry and red indicates a symmetry-broken state. a) For $\theta= 35^\circ$, the system is not in TRSB state at zero flux, and $\bar{\omega}_{pl}$ vanishes only together with the critical current. b) For $\theta =43^\circ$, $\bar{\omega}_{pl}$ vanishes at the field-driven phase transition, where $\bar{I}_c$ is finite.}
    \label{fig:plasma vs m}
\end{figure}

The soft-mode behavior can be also induced for fixed twist angle and temperature with an in-plane magnetic field. Using Eq. \eqref{eq:Fmag} and  \eqref{eq:joscur} we obtain the equations of motion for $\phi$ in the presence of a magnetic field, allowing us to compute $\bar{\omega}_{pl}(\bar{\Phi})$, shown in Fig. \ref{fig:plasma vs m}. For twist angles where the system is in TRSB phase at $\bar{\Phi} = 0$ (Fig. \ref{fig:plasma vs m}, (b)), a behavior similar to Fig. \ref{fig:plasma vs t,twist} is observed. Namely, $\bar{\omega}_{pl}(\bar{\Phi})$ vanishes at the symmetry-breaking transition for reasons described in Sec. \ref{sec:equil}. On the other hand, if the the system is not in the TRSB phase at $\bar{\Phi} = 0$ (Fig. \ref{fig:plasma vs m}, (a)), the plasma frequency vanishes only together with the critical current, and no symmetry-breaking transitions occur.

\red{Let us now discuss the implications of damping (finite $\beta_c$). In that case, the eigenfrequency of the linearized Eq. \eqref{Normalized RCSJ model} takes the form 
\begin{equation}
    \bar{\omega}_{0}(\theta,T) = 
    \pm
    \sqrt{\bar{\omega}^2_{pl}(\theta,T)-1/(4\beta_c(T))}
    +
    \frac{i}{2\sqrt{\beta_c(T)}}.
\end{equation}
Therefore, the resonant oscillation frequency is given by $|\text{Re}[\bar{\omega}_{0}(\theta,T)]|=\sqrt{\bar{\omega}^2_{pl}(\theta,T)-1/(4\beta_c(T)})$, which represents the plasma frequency for the damped system. Note that for $\bar{\omega}^2_{pl}(\theta,T)-1/(4\beta_c(T))<0$ the oscillations cease and become overdamped. Nonetheless, even in this regime signatures of TRSB state can be observed, see Sec. \ref{sec: reentrance}.
}

\subsubsection{Effects of explicit symmetry breaking}
\label{subsec:memory}

Experiments on twisted \bscco flakes \cite{Zhao2023} have shown an incomplete reversal of the Josephson diode effect -  ``memory effect" \cite{volkov_diode}, suggesting a different form of TRSB present in the system. The origin and magnitude of this explicit time reversal breaking term is presently unknown. Its effects can be modeled phenomenologically by adding a term $ -J_{m}\sin(\phi)$ to the free energy Eq. \eqref{free energy}. We will now demonstrate that the Josephson plasma frequency in presence of a DC bias current $I$ is sensitive to $J_m$, allowing its precise determination.

\begin{figure}
		\includegraphics[width=0.95\linewidth]{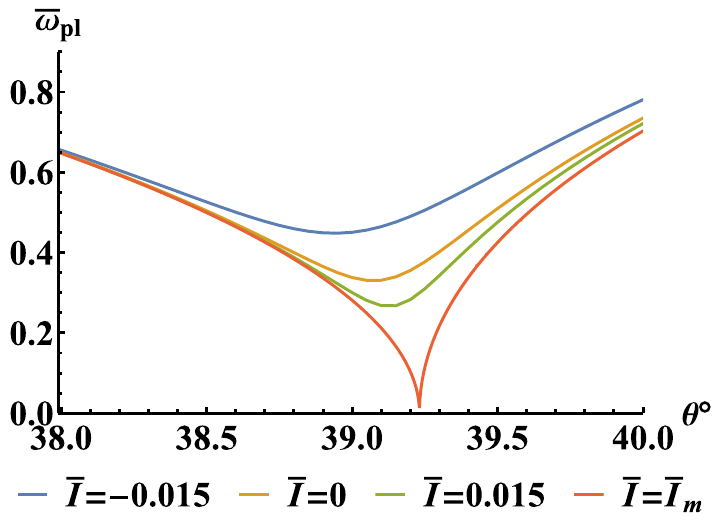}
	\caption{Josephson plasma frequency $\bar{\omega}_{pl}$ vs. $\theta$ with a finite explicit time-reversal breaking $\bar{J}_m=0.01$, varied applied DC current bias $\bar{I}$, and $\bar{J}^{\red{T=0}}_{c1}=10$ \red{at $T=0$}. We find that for $\bar{I}=\bar{I}_m=0.01$ the complete plasmon softening is recovered.
	}
	\label{fig:time reversal}
\end{figure}

In the presence of the explicit time reversal breaking term and the DC current $I$ (via adding $-I \phi$ to Eq.\eqref{free energy}), the plasma frequency is given by 
\begin{align}
    \bar{\omega}_{pl} = \Bigg(\bar{J}_{c1}(\theta\red{,T})\cos(\psi_0)-&\red{\left(1-\frac{T}{T_c}\right)^2}\cos(2\psi_0) \nonumber \\
    &-\bar{J}_m(\theta)\sin(\psi_0)\Bigg)^{1/2}
\end{align}
where $\psi_0$ is the new minimum of the free energy after the addition of the explicit time reversal breaking term $J_m$ and the DC current $I$, \red{and $\bar{J}_{m}=J_m/J_{c2}$}. In Fig. \ref{fig:time reversal}, we present the resulting $\bar{\omega}_{pl}(J_m, I, \theta\red{,T=0})$ for a fixed value of $J_m$ as a function of twist angle $\theta$. One notices that $\bar{\omega}_{pl}(J_m, I, \theta\red{,T=0})$ strongly depends on $I$. For $I=0$, one observes an incomplete softening of the Josephson plasmon, which is attributed to nonzero $J_m$ explicitly breaking the system's symmetry. However, finite $I$ can ``counteract" that effect, specifically leading to a complete softening for $I=J_m$. This allows us to determine the magnitude of $J_m$ precisely using soft-mode spectroscopy.


\subsection{Signatures of soft modes in linear impedance}
\label{sec:linimped}

We now discuss the implications of the existence of a soft mode, Sec. \ref{softmode} for experimental observables. The soft mode of the system can be probed by the linear voltage response of the system. Using the Josephson relations, the voltage through the junction is given by: 
\begin{equation}
    V(\tau\red{,T})=\left(\frac{
RJ_{c2}\red{(T)}
}{\sqrt{\beta_c\red{(T)}}}\right)\frac{d}{d\tau}\phi(\tau).
\end{equation}
The magnitude of the system's response is then characterized by the the impedance
\begin{equation}\label{impedance def}
    Z(\bar{\omega}\red{,T}) = V(\bar{\omega}\red{,T}) / I_0
\end{equation}
where $V(\bar{\omega})$ is the Fourier transform of $V(\tau)$.

We begin by discussing the linear response $V(\bar{\omega})$, which can be determined by solving \eqref{linear response} in the $\tau\rightarrow \infty$ limit (where the solution does not depend on the initial conditions). One finds 
$\phi_1(\tau\red{,\theta,T}) = \alpha_1\red{(\theta,T)}e^{i\bar{\omega}\tau}+\alpha_1^*\red{(\theta,T)}e^{-i\bar{\omega}\tau},$ where \red{$\alpha_1(\theta,T) = \bar{I}_0/[2(\text{Re}[\bar{\omega}_{0}(\theta,T)]^2-\bar{\omega}^2+i\bar{\omega}/\sqrt{\beta_c(T)}]$}.
From this, we find the linear impedance to be 
\begin{align}
    |Z_1(\bar{\omega}\red{,\theta,T})| &= \frac{RJ_{c2}\red{(T)}}{\sqrt{\beta_c\red{(T)}}}|\bar{Z}_1(\bar{\omega}\red{,\theta,T})| \nonumber \\
    &=\left|\frac{RJ_{c2}\red{(T)}}{\sqrt{\beta_c\red{(T)}}}\left(\frac{i\bar{\omega}}{2f_1(\red{\bar{\omega},}\theta\red{,T})}\right)\right|,
    \label{eq:z1}
\end{align}
where we introduced notation:
\red{
\begin{equation}
        f_n(\theta,\bar{\omega}\red{,T}) = \text{Re}[\bar{\omega}_{0}(\theta\red{,T})]^2-(n\bar{\omega})^2+i\frac{n\bar{\omega}}{\sqrt{\beta_c\red{(T)}}},
\end{equation}}
and $\bar{Z}_1(\bar{\omega}\red{,\theta,T}) =  \frac{\sqrt{\beta_c\red{(T)}}}{RJ_{c2}\red{(T)}} Z_1(\bar{\omega}\red{,\theta,T}) $.

\begin{figure}[h]
    \subfigure{
    \includegraphics[width=0.95\linewidth]{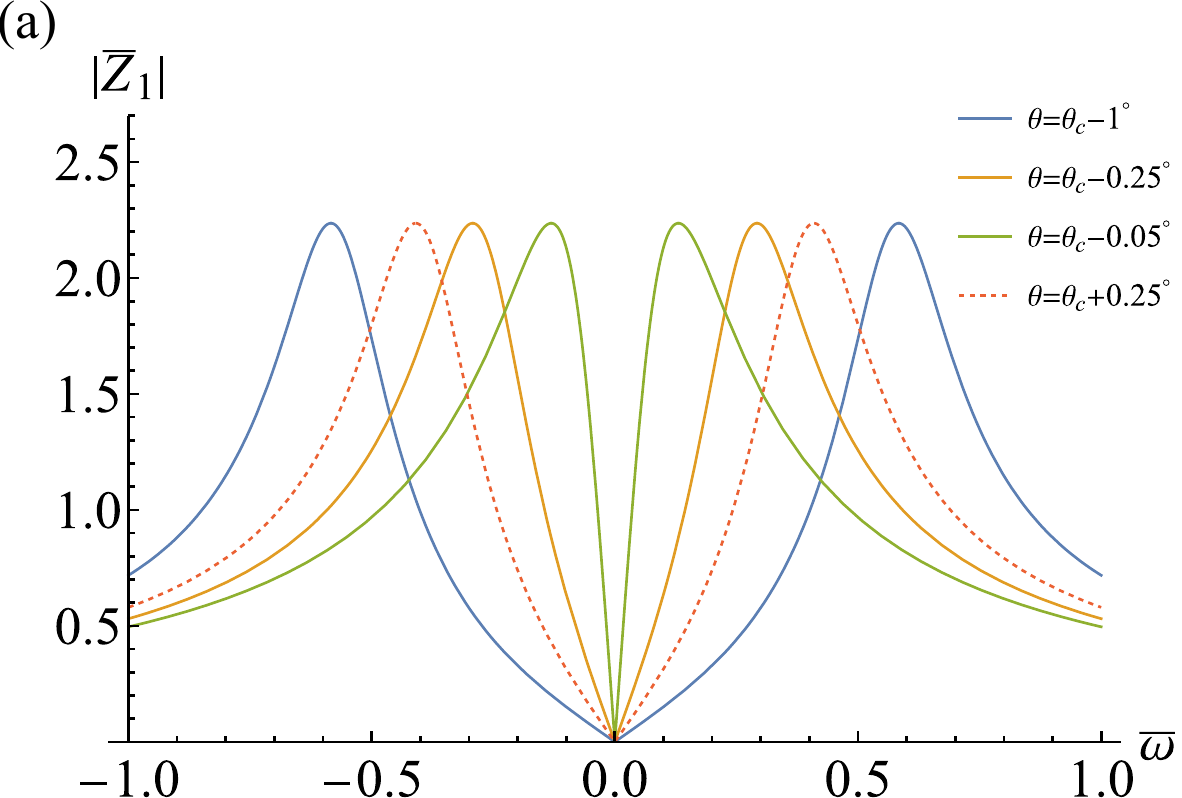}
    }
    \subfigure{
\includegraphics[width=0.95\linewidth]{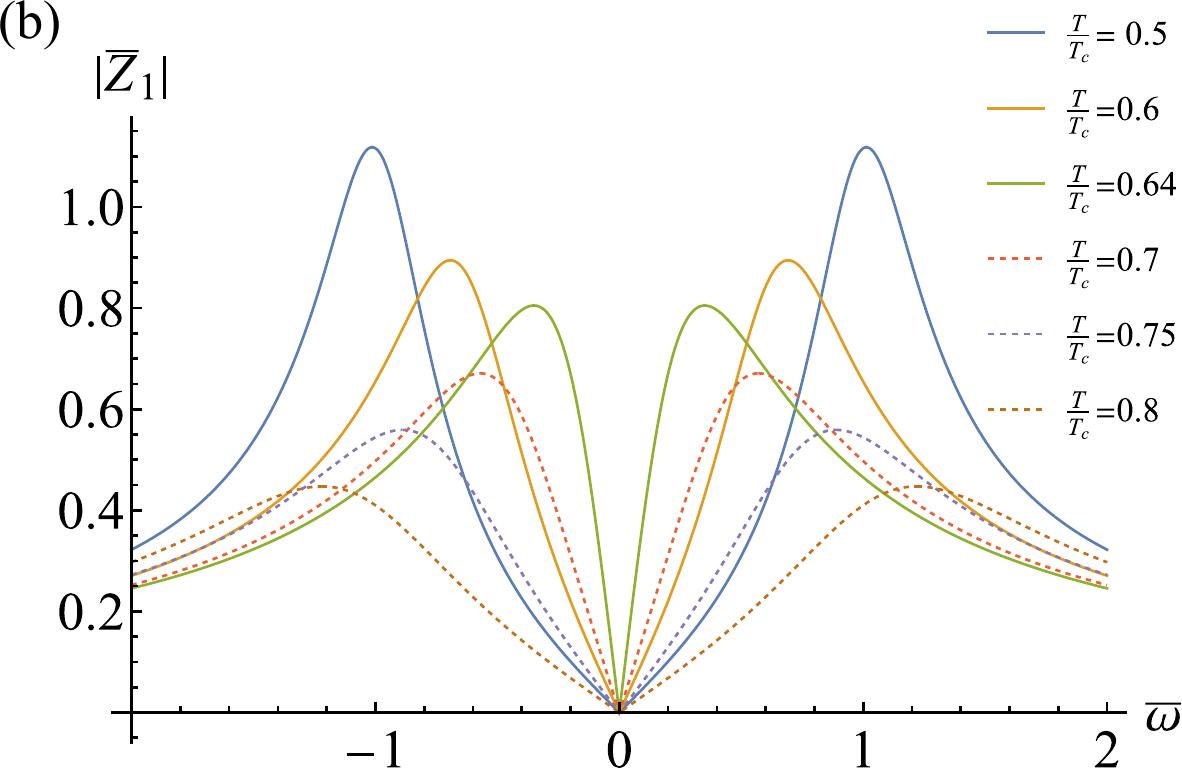}
    }
    \caption{Plot of the absolute values of the linear impedance $|\bar{Z}_1(\bar{\omega}\red{,\theta,T})|$ for a) different twist angles at $T=0$ and b) varying temperatures at $\theta=43^\circ$, where $T/T_c=0.65$ is the TRSB transition temperature. We've taken \red{$\bar{J}^{T=0}_{c1}=10$}, \red{yielding $\theta_{c\red{,T=0}}^+=39.2^\circ$}, and $\beta_c^{T=0}=20$. In both cases, the peak in the linear impedance shifts towards $\bar{\omega}= 0$ close to the transition indicating the presence of a soft mode.
    }
    \label{fig:Z^1}
\end{figure}

In Fig. \ref{fig:Z^1} we present the $|Z_1(\bar{\omega}\red
{,\theta,T})|$ as a function of twist angle (a) and temperature (b) for $\beta_c\red{(T=0)}=20$. For the first case, the only varying parameter is $\bar{J}_{c1}(\theta\red{,T=0}) = \bar{J}^{\red{T=0}}_{c1} \cos(2 \theta)$, while for the temperature variation $\beta_c(T) = \beta_c^{T=0} (1-T/T_c)^2$. In both cases, there is a noticeable peak in $|\bar{Z}_1(\bar{\omega}\red{,\theta,T})|$ that shifts towards zero frequency on going through the transition. We note that right at the transition, where $\bar{\omega}_{pl}^2(\theta\red{,T}) =0$, nonlinear effects are expected to be important at low frequencies, which will be discussed below.

\section{Nonlinear response: second harmonic generation in the TRSB phase}\label{sec: reentrance}

In this section we consider nonlinear responses of the twisted superconductor interface to a harmonic drive. As the system approaches phase transition, the flattening of the free energy, reflected in $\omega_{pl}$ going to zero, suggests increased importance of nonlinear response and anharmonic effects. To capture the nonlinear response of the system, we can examine the higher order responses to the harmonic drive $I_0(t)$. As the $n$-th order of the perturbative solution for the phase are to the $n$th power of the driving current, $\bar{I}_0$, we will define the nonlinear impedance as 
\begin{equation}
    Z_n(\bar{\omega}\red{,\theta,T})=V_n(n\bar{\omega}\red{,\theta,T})/\bar{I}_0^n\red{(T)}.
\end{equation}
We will similarly define a dimensionless nonlinear impedance as $\bar{Z}_n(\bar{\omega}\red{,\theta,T}) = Z_n(\bar{\omega}\red{,\theta,T})\left(\frac{
RJ_{c2}\red{(T)}
}{\sqrt{\beta_c\red{(T)}}}\right)^{-1}$


\begin{figure}[h]
    \subfigure{\includegraphics[width=0.95\linewidth]{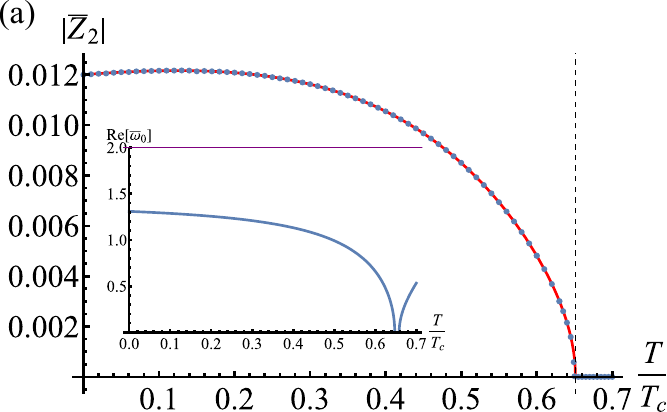}}
    \subfigure{\includegraphics[width=0.95\linewidth]{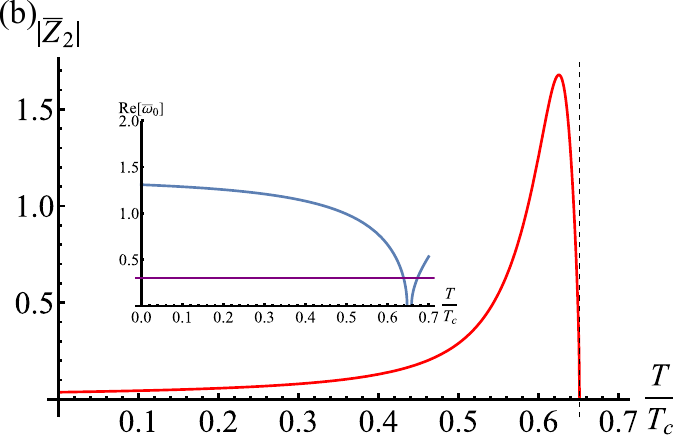}}
    \caption{Plot of $\bar{Z}_2$ vs. $T/T_c$ for $\beta_c\red{(T=0)}=6$, \red{$\bar{J}^{T=0}_{c1}=10$, $\theta_{c,T=0}^+=39.2^\circ$, and  $\theta=43^\circ$}. We choose \red{a) $\bar{\omega}=2$ as $|\text{Re}[\bar{\omega}_{0}^{T=0}]|=1.3$ and b) $\bar{\omega}=0.3$ to demonstrate the effect of resonance. The insets correspond to $|\text{Re}[\bar{\omega}_0]|$ vs. $T/T_c$, where the purple horizontal line indicates the driving frequency.} Blue points are obtained from numerical solution of Eq. \eqref{Normalized RCSJ model} (see also Sec. \ref{sec:non-eq}) and red is the analytical result, Eq. \eqref{eq:2ndan}. The vertical dashed line corresponds to the $T/T_c$ value for which the system undergoes the phase transition.}
	\label{fig:second order impedance}
\end{figure}

We begin by discussing the second harmonic response $Z_2$. As we can see from \eqref{2nd harm response}, the second harmonic term is ``driven" by the first via the term proportional to $b$. The steady-state second harmonic component of second-order phase difference is $\phi_2^{2\bar{\omega}}(\tau\red{,\theta,T}) = \beta_1\red{(\theta,T)} e^{i2\bar{\omega} \tau}+\beta_1^*\red{(\theta,T)} e^{-i2\bar{\omega} \tau}$ where $\beta_1\red{(\theta,T)} = -\alpha_1^2b(\theta\red{,T})/f_2(\theta\red{,\bar{\omega},T})$. Note that there is also a $\bar{\omega}=0$ component of $\phi$ at second order, but it does not lead to a contribution to voltage. The second harmonic impedance is thus 
\begin{equation}
    \bar{Z}_2(\bar{\omega}\red{,\theta,T})= \frac{-i b(\theta\red{,T})\bar{\omega}}{2f_1^2(\theta,\bar{\omega}\red{,T})f_2(\theta,\bar{\omega}\red{,T})}.
    \label{eq:2ndan}
\end{equation}
The results are shown in Fig. \ref{fig:second order impedance}, as a function of temperature for a fixed $\bar{\omega} = 2$. The second harmonic vanishes identically in the symmetric phase, while it is nonzero in the TRSB phase. This can be tracked down to $b(\theta, \red{T,\phi=0})=0$ in the symmetric phase. In the time reversal symmetric case, the free energy is symmetric about $\phi_0=0$, resulting in only even powers in the Taylor expansion in $\phi-\phi_0$. However, an odd power is required to produce a second harmonic response.

One can conclude that second harmonic response is a {\it necessary and sufficient} criterion for TRSB state, unlike the diode effect \cite{volkov_diode}. There is, however one exception: at $\theta=45^\circ$ the free energy is symmetric around  $\phi_0=\pm\pi/2$ and $b(\theta,\red{T,}\pm \pi/2) = 0$ resulting in zero second harmonic response (the diode effect has also been predicted to vanish exactly at 45$^\circ$ \cite{volkov_diode}). However,in contrast to the diode effect, second harmonic response is present at all other twist angles within the TRSB phase, regardless of the $\beta_c$ value. \red{We also would like to comment that our proposal does not distinguish between different microscopic mechanisms of TRSB. Indeed, both disorder \cite{yuan2023,yuan2024} and higher-order tunneling processes \cite{volkov_2025,Can2021} lead to the same form of Eq.\eqref{free energy}.}

We have also considered perturbatively the higher-harmonic responses. They do not bring additional qualitative information on the phase transition and are discussed in Appendix \ref{app:3rd}.

\section{Strong nonlinearity and nonequilibrium phase transitions}
\label{sec:non-eq}

\begin{figure}
    \centering
    \includegraphics[width=0.95\linewidth]{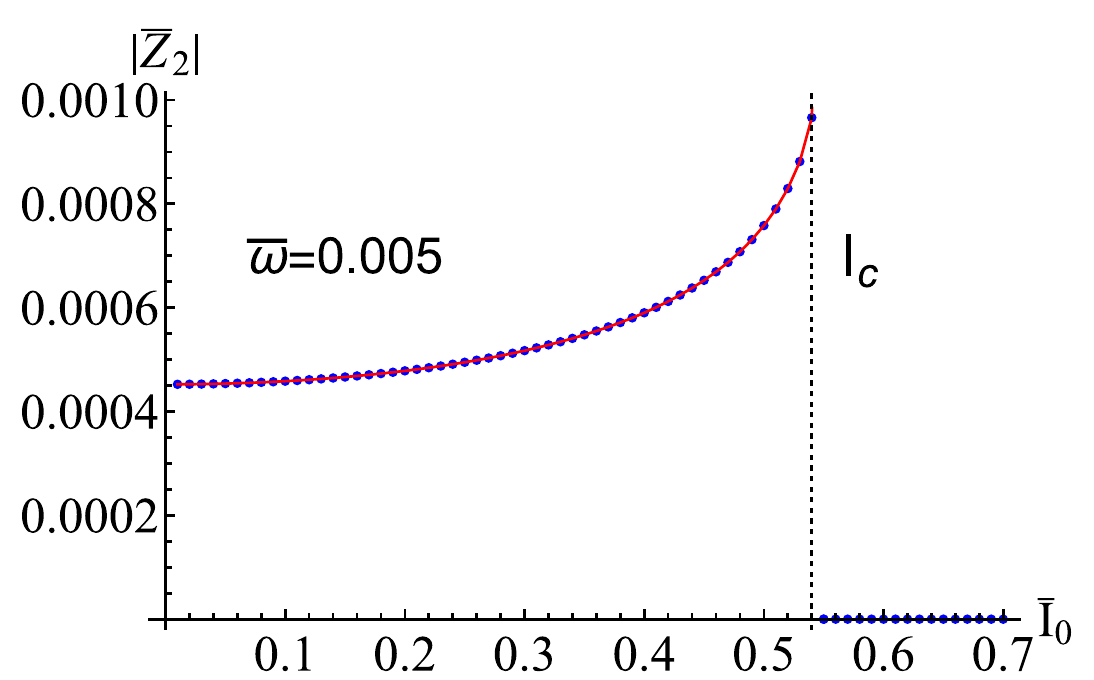}
    \caption{Second harmonic impedance a function of AC drive amplitude $\bar{I}_0$ for $\bar{\omega} =0.005 \red{\ll |\text{Re}[\bar{\omega}^{T=0}_{0}]|=1.3}$, $\beta_c^{T=0}=6$,  \red{$\bar{J}^{T=0}_{c1}=10$, $\theta=43^\circ$, $\theta_{c,T=0}^+=39.2^\circ$, and $T=0$}. Blue points are obtained from the numerical solution of the RCSJ equation with $\beta_c^{T=0}=6$ \eqref{Normalized RCSJ model}. The red curve is obtained numerically from the quasi-adiabatic approximation (see text).}
    \label{fig:I <I_c second harm}
\end{figure}

\begin{figure*}[t]
    \subfigure{\includegraphics[width=0.325\textwidth]{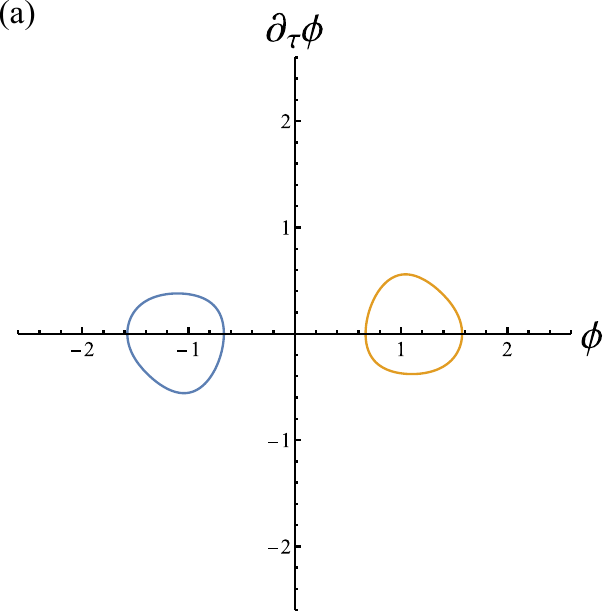}}
    \subfigure{\includegraphics[width=0.325\textwidth]{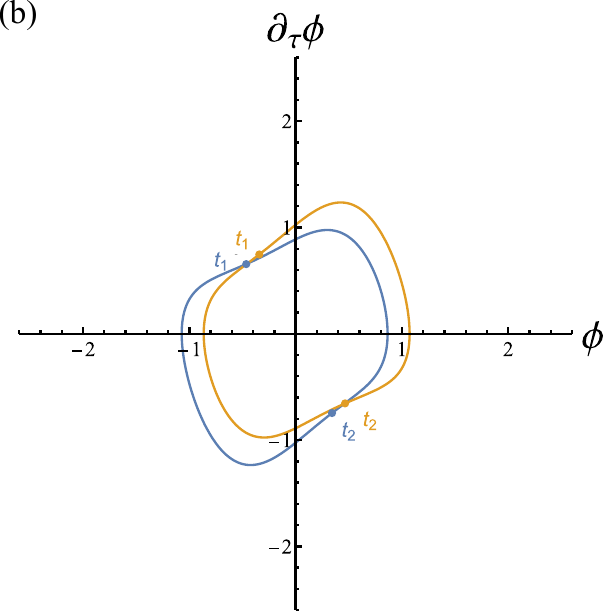}}
    \subfigure{\includegraphics[width=0.325\textwidth]{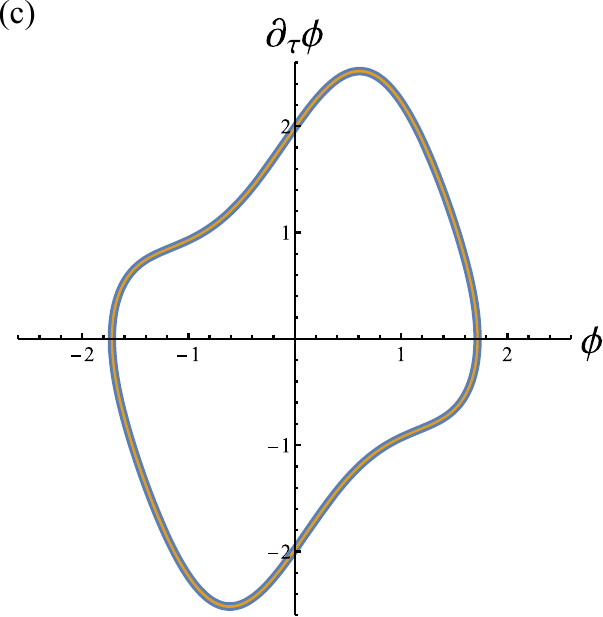}}
    \subfigure{\includegraphics[width=0.325\textwidth]{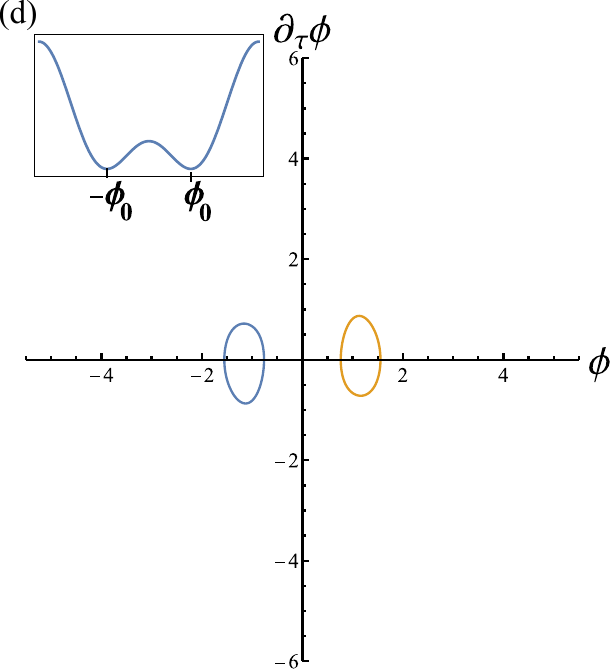}}
    \subfigure{\includegraphics[width=0.325\textwidth]{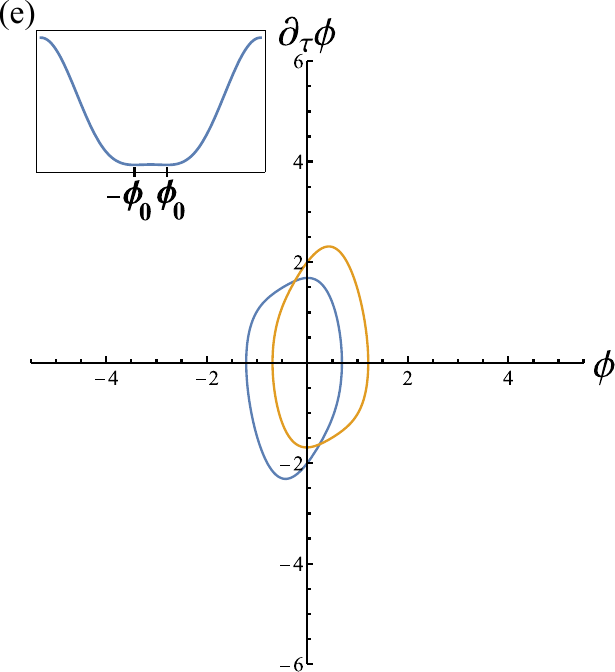}}
    \subfigure{\includegraphics[width=0.325\textwidth]{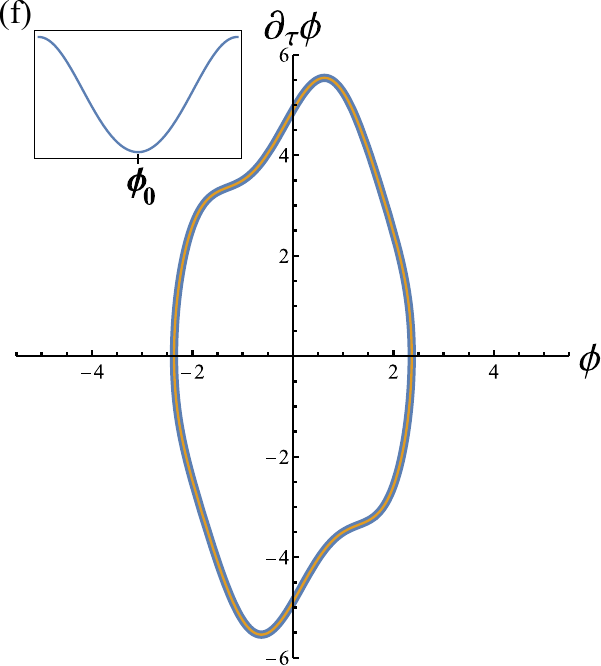}}
    \caption{Phase portraits of Eq. \eqref{Normalized RCSJ model} in the steady state at intermediate frequencies (first row $\bar{\omega}=1$, second row $\bar{\omega}=2$ with $|\text{Re}[\bar{\omega}^{T=0}_{0}]|=1.3$), where the insets correspond to the time averaged free energy, Eq. \eqref{eq:Fav}. The blue and orange curve correspond to different initial conditions $\phi(0)=\pm\phi_0, \partial_\tau \phi(0)= 0$.  The apparent intersection of solutions in (b,e) in fact corresponds to different times - the points in b) show the two solutions at the same time $t_1$ and $t_2$. For large enough drive, however, he steady state solutions overlap completely (c,f). The driving current values are: a) 
    $\bar{I}_0=0.8$, b) $\bar{I}_0=1.147$, and c) $\bar{I}_0=2$, d) 
    $\bar{I}_0=1$, e) $\bar{I}_0=2$, and f) $\bar{I}_0=5$. For all panels, $T=0$, $\bar{J}^{T=0}_{c1}=10$, $\theta=43^\circ$, and $\theta_{c,T=0}^+=39.2^\circ$ was taken.}
    \label{phase_portrait}

\end{figure*}

We now demonstrate that strong AC driving leads to non-equilibrium phase transitions in twisted superconductor interfaces. \pv{We follow previous works on driven ferromagnets \cite{dyntr1} and Josephson junctions \cite{dyntr2} in identifying such transitions as the emergence or disappearance of a nonzero time-averaged value of $\langle\phi(\tau)\rangle_\tau$, which, in the steady-state, can have several values related by the system's symmetry. In \cite{dyntr2}, it has been also shown that such transitions are accompanied by the emergence of a second-harmonic response.} To go beyond the perturbative treatment in Sec. \ref{sec: reentrance} above, we study Eq. \ref{Normalized RCSJ model} numerically. To determine the output voltage, we computed the solution of $\phi(\tau)$ at sufficiently long time, such that the Fourier transform $V(\omega)$ is not affected by the initial conditions (see Appendix \ref{sec:app_num}). For low values of $\bar{I}_0 = 0.01$, this procedure reproduces the perturbative results for $Z_2(\omega)$, see Fig. \ref{fig:second order impedance}. 

In Fig. \ref{fig:I <I_c second harm}, we present the numerical result for $\overline{Z}_2(\bar{\omega}=0.005)$ as a function of the driving current amplitude. $\overline{Z}_2$ increases up to a critical value of $\bar{I}_0$, after which it abruptly drops to zero and remains so for a finite range of $\bar{I}_0$. In the low-frequency limit (where $\phi$ adiabatically tracks the position of the local minimum), this can be understood as follows. If in the absence of the drive the system has two equilibrium states, $\phi=\pm \phi_0$. A weak drive will only induce small oscillations of $\phi$ around each of the two minima{, with a nonzero time-averaged $\phi$ value}. $\bar{I}_0^{cr}$ marks the point where one of the two minima ceases to exist at a certain point in time. In that case, $\phi$ will oscillate between the two minima of the potential, such that the motion in the first half-period and second half-period are related by $\phi(t+\red{\mathcal{T}}/2) = -\phi(t)$, $\red{\mathcal{T}}=2\pi/\omega$, which leads to the vanishing second harmonic response $\phi(2 \omega) = 0$ \pv{as well as time-averaged $\langle\phi(\tau)\rangle_\tau=0$}. Thus, $\bar{I}_0^{cr}$ corresponds to a bifurcation of the dynamics of $\phi(t)$: whereas for $\bar{I}_0<\bar{I}_0^{cr}$ there are two asymptotic solutions (limit cycles), oscillating around $\phi_0$ and $-\phi_0$, for $\bar{I}_0>\bar{I}_0^{cr}$ there is only one solution type oscillating between two minima and thus dynamically "restoring" the symmetry. The red line in Fig. \ref{fig:I <I_c second harm} corresponds to the second harmonic impedance calculated for $\phi(\tau) = \phi_0(I(\tau))$, i.e. using the adiabatic approximation explained above and shows good agreement with the numerical calculation.

\pv{The dynamical phase transition discussed above persists for larger driving frequencies.  In Fig. \ref{phase_portrait}, we show the phase portraits of Eq. \eqref{Normalized RCSJ model} for driving frequency $\overline{\omega}$ of the order of ${\rm Re}[\overline{\omega}_0]$ on increasing the driving amplitude (from left to right. For sufficiently large amplitude the two solutions obtained with initial conditions in the two distinct equilibria lead to identical steady-state dynamics with zero average $\langle\phi(\tau)\rangle_\tau$ (c,f) - indicating a dynamical phase transition into a symmetric phase.}

\begin{figure*}[t]
    \subfigure{\includegraphics[width=0.325\textwidth]{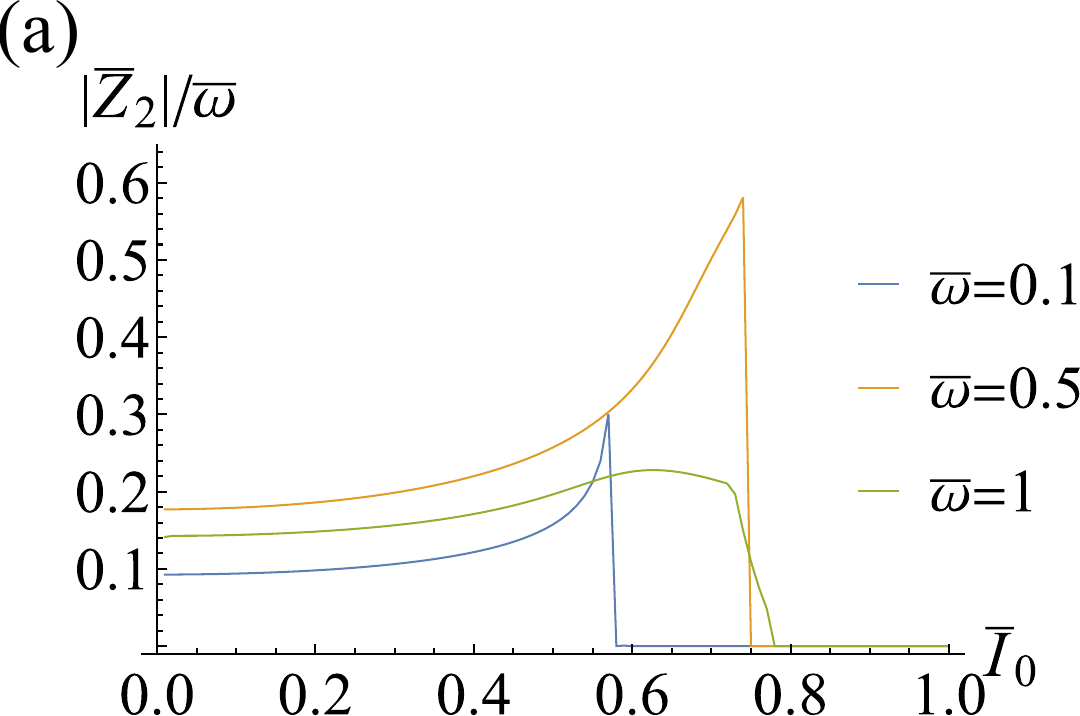}}
    \subfigure{\includegraphics[width=0.325\textwidth]{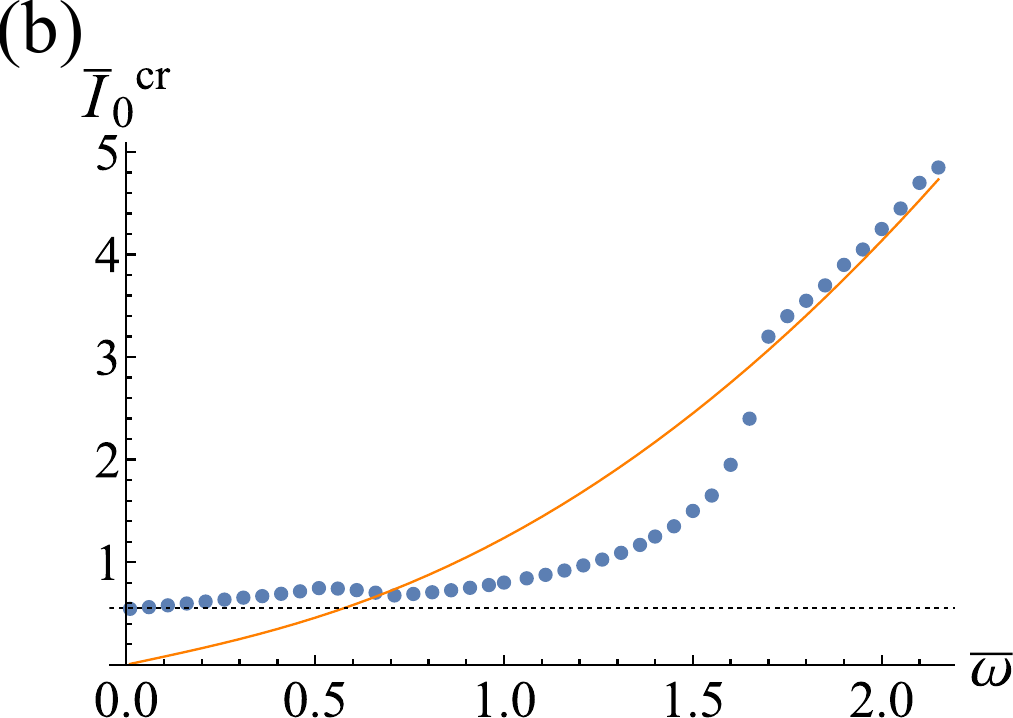}}
    \subfigure{\includegraphics[width=0.325\textwidth]{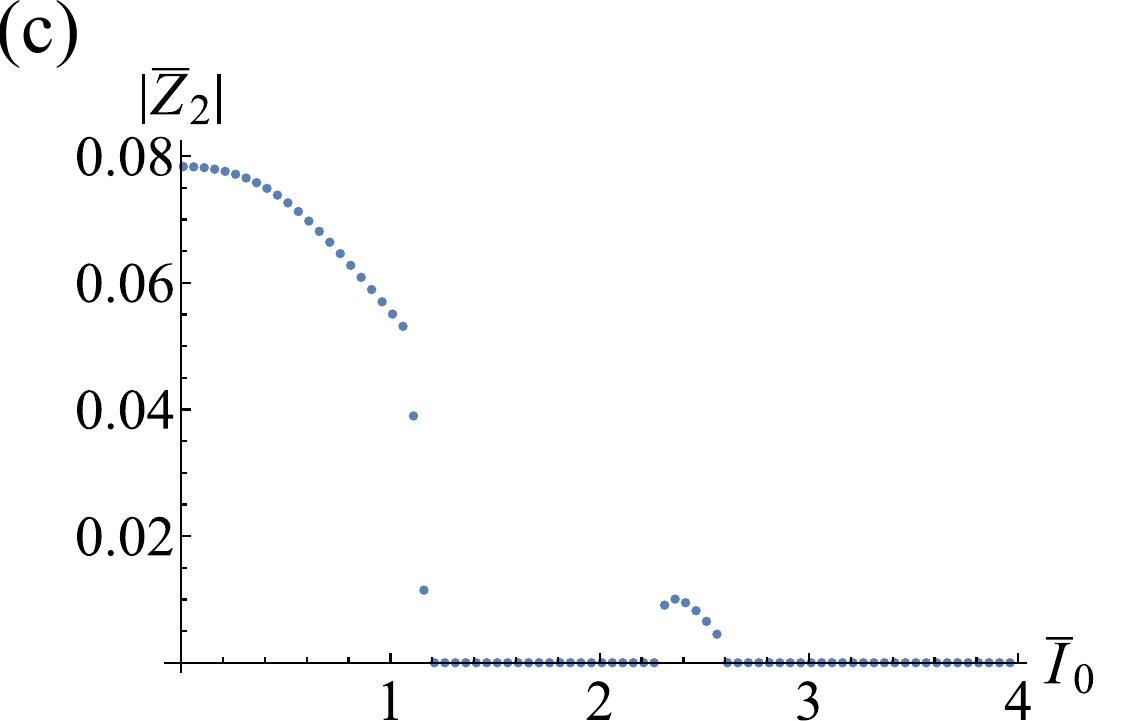}}
    \caption{Second-harmonic response for varying $\bar{I}_0$ and $\bar{\omega}$ in the under-damped limit.  (a) Second harmonic impedance \red{normalized by $\bar{\omega}$} as a function of AC drive amplitude $\bar{I}_0$ for varying $\bar{\omega}$. (b) The critical current for the disappearance of the second harmonic plotted against the driving frequency of the bias current. The oragne graph is the predictions from the high frequency results \eqref{eq:ic0cran}, and the blue circles correspond to the numerical results. The dotted line is the prediction from the low frequency case. (c) $|\bar{Z}_2(\bar{I}_0)|$ at $\bar{\omega}=1.4$ for an extended  $\bar{I}_0=2$ range. We see the reentrance of second harmonic around $\bar{I}_0=2$. For all panels, \red{$T=0$, $\bar{J}^{T=0}_{c1}=10$, $\theta=43^\circ$, $\theta_{c,T=0}^+=39.2^\circ$ and $\beta^{T=0}_c=1.5$} was taken, \red{resulting in $|\text{Re}[\bar{\omega}_0]|=1.26$ and $\text{Im}[\bar{\omega}_0]=0.4$.}}
        \label{fig:nonperturb numerics}
\end{figure*}

\begin{figure}[h!]
    \centering
    \includegraphics[width=0.8\linewidth]{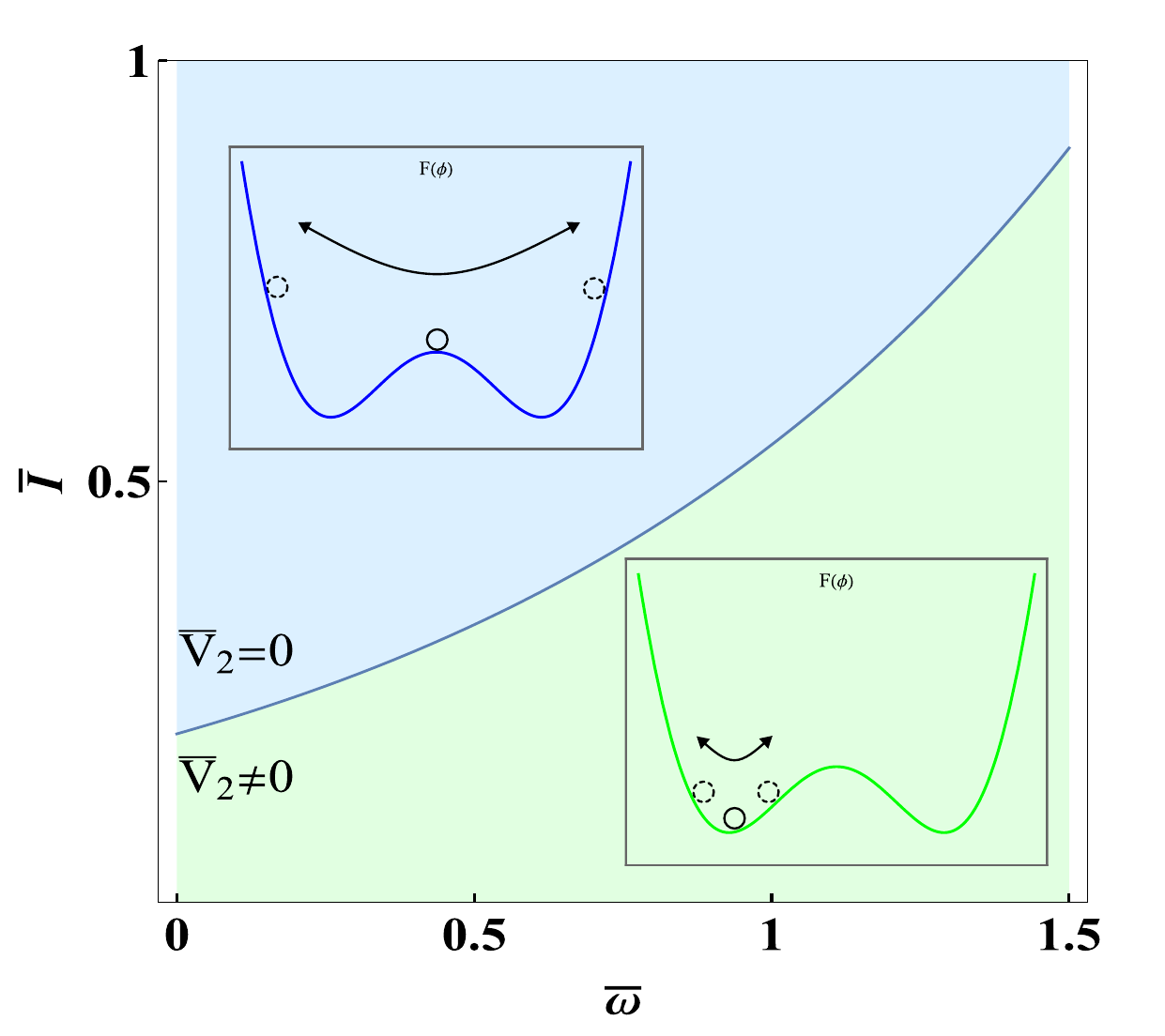}
    \caption{Schematic phase diagram of the interface for varied driving frequency and driving current, without accounting for reentrant behavior. We begin with a system exhibiting TRSB \red{in the overdamped case without resonance}, so for small driving current, second harmonic is generated. We find that above $\bar{I}_c(\theta,\beta_c)$ (blue line), the system becomes TRS as indicated by the loss of second harmonic (Sec. \ref{sec: reentrance}).}
    \label{fig:phase-diagram2}
\end{figure}

Remarkably, the abrupt vanishing of the second harmonic voltage on increasing $I_0$ also persists for any driving frequency $\bar{\omega}$, as shown in Fig. \ref{fig:nonperturb numerics} (a). One can then define the critical $\bar{I}_0^{cr}(\bar{\omega})$, where the second harmonic voltage first vanishes, shown in Fig. \ref{fig:nonperturb numerics} (b). $\bar{I}_0^{cr}(\bar{\omega})$ therefore separates two dynamical phases, characterized by \red{the existence of one or two possible steady-state solutions, shown in Fig. \ref{phase_portrait}, and the absence or presence of the second harmonic voltage}. We illustrate the tentative phase diagram of the system in Fig. \ref{fig:phase-diagram2} \red{for the overdamped case, which we further discuss below}. Thus, we demonstrate that AC current can drive non-equilibrium phase transitions in twisted superconductors.

\subsection{Reentrant dynamical transitions}

Increasing the current amplitude even further beyond $\bar{I}_0^{cr}$, Fig. \ref{fig:nonperturb numerics} (c), one observes a reentrant behavior for $Z_2$, indicating a dynamical reentrant phase transition.

A full analytical understanding of this phenomenon can be obtained in the high-frequency limit $\bar{\omega}^2+\bar{\omega}/\sqrt{\beta_c\red{(T)}}\gg  1$. In this case, the Josephson (nonlinear) terms in Eq. \eqref{Normalized RCSJ model} can be considered perturbatively \footnote{Specifically, the Josephson current $\bar{I}_J$ can be bound by $\min\left[\bar{I}_c (\phi-\phi_0),\bar{I}_c\right]$. Neglecting $\bar{I}_c$ one gets $(\phi-\phi_0)_{max} \sim \bar{I}_0/(\bar{\omega}^2+\bar{\omega}/\sqrt{\beta_c})$, such that (taking $\bar{I}_c\sim 1$ for $\theta$ around 45$^\circ$), $I_J\ll I_0$ for all $I_0$.}. One then expects the solution to take the form $\red{\phi}(\tau) = \red{\phi}_0- a \sin(\bar{\omega}\tau + \psi_0) + \delta\red{\phi}_{n>1}(\tau)$. Neglecting the Josephson terms, one gets $a = \frac{\sqrt{\red{\beta_c(T)}} I_0}{\bar{\omega}\sqrt{\red{\beta_c(T)}\bar{\omega}^2+1} }$ and $\psi_0 = \arccos \sqrt{
\frac{\red{\beta_c} \bar{\omega}^2}{\red{\beta_c} \bar{\omega}^2+1}
}$. For the Josephson terms we can then use the Jacobi-Anger expansion:
\begin{equation}
\label{eq:jacobi-anger}
\begin{gathered}
    \sin[n\left\{a \sin(\bar{\omega}\tau + \psi_0)+\red{\phi}_0\right\}] = 
\\
=2\cos[n \red{\phi}_0] 
\sum_{k=1}^\infty 
J_{2n-1}(na)
\sin[(2k-1)(\bar{\omega}\tau+\psi_0)]
\\
+
\sin[n \red{\phi}_0] 
[J_0(na)+2 \sum_{k=1}^\infty J_{2 k}(na) \cos[2k(\bar{\omega}\tau+\psi_0)]].
\end{gathered}
\end{equation}
The expansion can be then reintroduced into the equations of motion and solved for each harmonic. However, the 0th harmonic is special because neither the derivative terms nor drive contribute, therefore it has to vanish exactly leading to:
\begin{align}
    0= \bar{J}_{c1}(\theta\red{,T}) J_0&\left[\frac{\sqrt{\red
    {\beta_c(
T)}} I_0}{\bar{\omega}\sqrt{\red{\beta_c(T)}\bar{\omega}^2+1}}\right]
    \sin[\red{\phi}_0] \nonumber \\
    &-
    J_0\left[2 \frac{\sqrt{\red{\beta_c(T)}} I_0}{\bar{\omega}\sqrt{\red{\beta_c(T)}\bar{\omega}^2+1}}\right]
    \sin[2 \red{\phi}_0]
    \label{eq:besselphi0}
\end{align}
where $J_n(x)$ is the Bessel function of first kind. This equations admit a trivial solution $\red{\phi}_0=\pi n$ and a nontrivial one:
\begin{equation}
    \red{\phi}_0' = \arccos 
    \left\{
    \frac{\bar{J}_{c1}(\theta\red{,T})} {2} \frac{J_0\left[\frac{\sqrt{\red{\beta_c(T)}} I_0}{\bar{\omega}\sqrt{\red{\beta_c(T)}\bar{\omega}^2+1}}\right]}
    {J_0\left[2 \frac{\sqrt{\red{\beta_c(T)}} I_0}{\bar{\omega}\sqrt{\red{\beta_c(T)}\bar{\omega}^2+1}}\right]}
    \right\}.
    \label{eq:besselphi0}
\end{equation}
We note that for the trivial solution, even harmonics of $\red{\phi}(\tau)$ are all zero, being absent from the Josephson term expansion, Eq. \eqref{eq:jacobi-anger}. \red{We can obtain the same result from the time averaged effective free energy, Eq. \eqref{free energy} with $\red{\phi}(\tau) = \red{\phi}_0- a \sin(\bar{\omega}\tau + \psi_0)$. The result takes the form:
\begin{align}
    \Big\langle F(\phi(t))\Big\rangle_{t}  =  &-\bar{J}_{c1}(\theta,T) J_0\left[\frac{\sqrt{\red{\beta_c(T)}}I_0}{\bar{\omega}\sqrt{\red{\beta_c(T)}\bar{\omega}^2}+1}\right]\cos \left(\phi_0\right) \nonumber \\ 
    &+\frac{1}{2} J_0
    \left[2\frac{\sqrt{\red{\beta_c(T)}}I_0}{\bar{\omega}\sqrt{\red{\beta_c(T)}\bar{\omega}^2}+1}\right] \cos \left(2 \phi_0\right).
    \label{eq:Fav}
\end{align}
Minimizing $ \Big\langle F(\phi(t))\Big\rangle_{t}$ over $\phi_0$ results in Eq. \eqref{eq:besselphi0}. In Fig.
\ref{phase_portrait} we show  $    \Big\langle F(\phi(t))\Big\rangle_{t}[\phi_0]$ in the insets; remarkably, the change in the shape of  $    \Big\langle F(\phi(t))\Big\rangle_{t}[\phi_0]$ from double-minimum to single-minimum coincides with the dynamical phase transition. This allows to interpret the AC-current induced phase transition in terms of the renormalized, time-averaged free energy Eq. \eqref{eq:Fav}.
}

We note that when $\red{\phi}_0'$ is the stable solution, $-\red{\phi}_0'$ is also a solution, such that the dynamical system has two limit cycles, \red{which can be seen in Fig. \ref{phase_portrait}}. This result confirms the nature of the dynamical transition discussed in the adiabatic limit above.

Eq. \eqref{eq:besselphi0} further allow us to quantitatively determine the phase boundaries of the dynamical symmetry-breaking phase. Indeed, $\red{\phi}_0'$ solution only exists when the argument of the $\arccos$ is less than $1$. Specifically, it first ceases to exist on increasing $I_0$ when:
\begin{equation}
\bar{J}_{c1}(\theta\red{,T}) J_0\left[\frac{\sqrt{\red{\beta_c(T)}} I_0^{cr}}{\bar{\omega}\sqrt{\red{\beta_c(T)}\bar{\omega}^2+1}}\right]=2J_0\left[2 \frac{\sqrt{\red{\beta_c(T)}} I_0^{cr}}{\bar{\omega}\sqrt{\red{\beta_c(T)}\bar{\omega}^2+1}}\right],
\label{eq:ic0cran}
\end{equation}
which is an equation determining $I_0^{cr}$. In Fig. \ref{fig:nonperturb numerics} (b, yellow line) we show that this analytical result is in excellent agreement with numerical data already above $\bar{\omega}\sim 1$ for $\beta_c\sim 1$. Due to the oscillatory character of Bessel functions in Eq. \eqref{eq:besselphi0}, increasing $I_0$ further can lead to a reentrant appearance of the nontrivial solution, explaining the behavior in Fig. \ref{fig:nonperturb numerics} (c).  

Eq. \eqref{eq:besselphi0} also implies that AC driving can be used to control Josephson junctions in the conventional regime, $\bar{J}_{c1}\red
{(\theta,T)}\ll 1$. In that case, nontrivial solutions will appear in the vicinity of the zeroes of $J_0 \left[\frac{\sqrt{\red{\beta_c(T)}} I_0}{\bar{\omega}\sqrt{\red{\beta_c(T)}\bar{\omega}^2+1}}\right]$. Unlike the case of tuning by magnetic field considered in Sec. \ref{sec:equil}, the vanishing of the first term in Eq. \eqref{eq:besselphi0} does not imply vanishing of the second one; in fact $J_0(2 x_n)<0$, where $J_0(x_n)=0$. Therefore, tuning $I_0$ to suppress the first harmonic will effectively make the second harmonic dominant, although with a different sign from its original one. Thus, for $J_{c2}\red{(T)}<0$ a bistable state similar to the TRSB one will appear around $\frac{\sqrt{\red{\beta_c(T)}} I_0}{\bar{\omega}\sqrt{\red{\beta_c(T)}\bar{\omega}^2+1}} = x_n$, while for $J_{c2}\red{(T)}>0$ a switching between $0-$ and $\pi-$ junction regime will occur at this driving amplitude. 

\begin{figure}[!tp]
    \centering
    \subfigure{\includegraphics[width=0.9\linewidth]{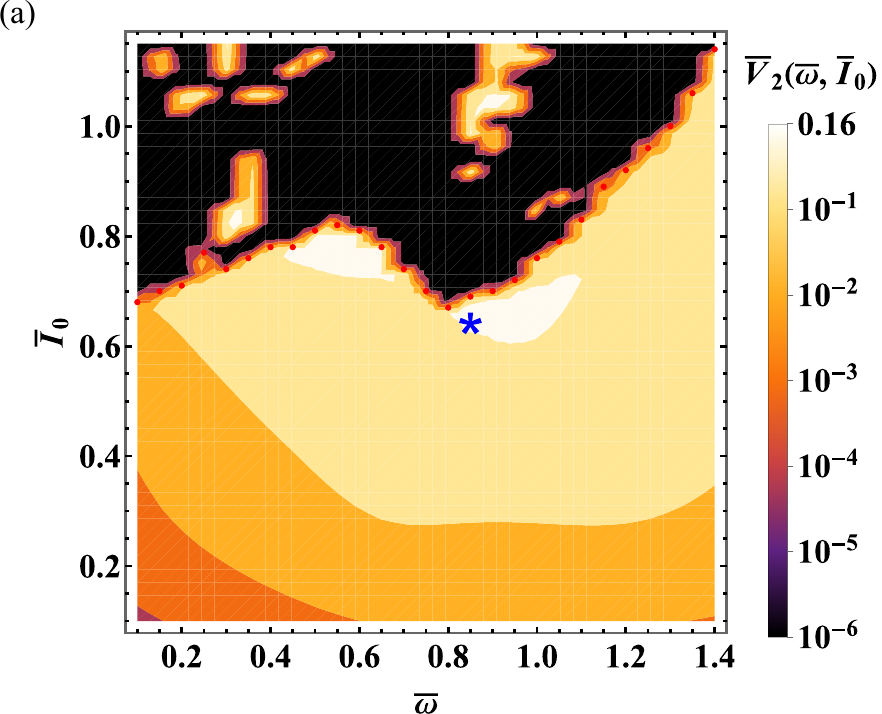}}
    \subfigure{\includegraphics[width=0.9\linewidth]{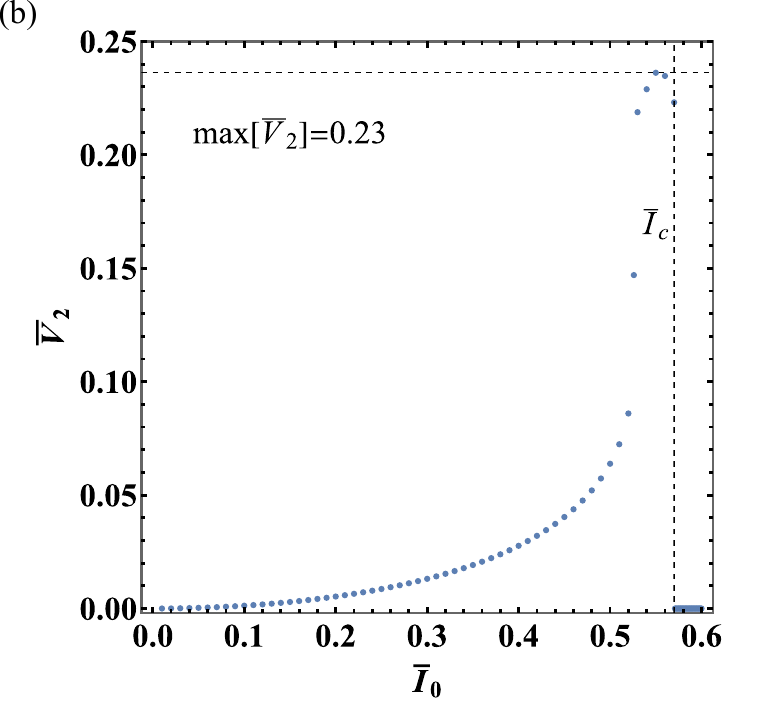}}
    \subfigure{\includegraphics[width=0.9\linewidth]{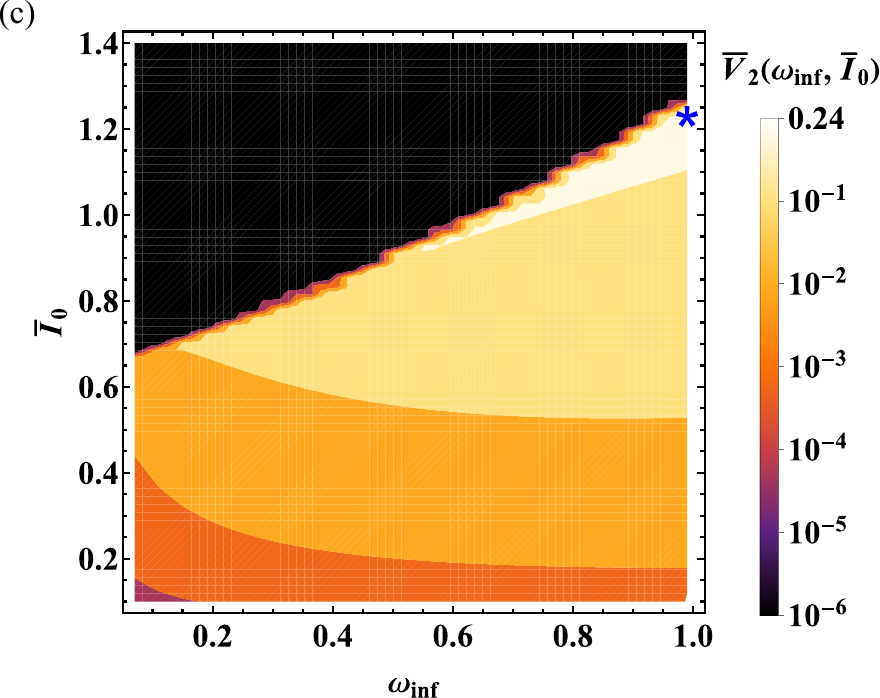}}
        \caption{Maps of the steady-state second harmonic voltage $\bar{V}_2$ as a function of $\bar{I}_0$ and $\bar{\omega}$ for \red{$T=0$, $\bar{J}^{T=0}_{c1}=15$, $\theta=44^\circ$, which is within TRSB phase in equilibrium ($\theta>\theta_{c,T=0}^+=41.2^\circ$)}, (a) $\beta^{\red{T=0}}_c=2$ \red{($|\text{Re}[\bar{\omega}_0]|=1.32$, $\text{Im}[\bar{\omega}_0]=0.35$)}, (b) $\bar{\omega} = 0.55$, $\beta^{\red{T=0}}_c=25$ \red{($|\text{Re}[\bar{\omega}_0]|=1.36$, $\text{Im}[\bar{\omega}_0]=0.1$)}, and (c) $\beta_c\rightarrow 0$ (note that we use a different frequency normalization $\omega_{\text{inf}}=\omega(2e J_{c2}\red{(T)}R /\hbar)^{-1}$ in this case). The red dots in the first contour plot indicate the critical AC current $\bar{I}_0^{cr}$ beyond which $\bar{V}_2$ vanishes for the first time. The blue star in both contour plots indicate the point with the largest $\bar{V}_2$ value in the plot. Finally, the vertical dashed line in the (b) corresponds to the critical AC current of $\bar{I}_0^{cr} = 0.55$.}
    \label{fig:max V2}
\end{figure}

To explore the reentrant phases beyond the high-frequency approximation, we have evaluated $\overline{V}_2$ numerically in the low-frequency regime, Fig. \ref{fig:max V2}. One notes that reentrant phases appear as ``islands" in the phase diagram for $\beta_c\red{(T=0)}=2$ (Fig. \ref{fig:max V2}, (a)), but are absent (for the $\bar{I}_0$ values shown) in the $\beta_c \to 0$ (overdamped junction) case (Fig. \ref{fig:max V2}, (c)). In all the cases, however, one notices that the maximal $V_2$ value as a function of $I_0$ and $\omega$ is similar, around $(0.05-0.25) J_{c2}^{\red{T=0}} R$. \red{Further, we find that the second harmonic persists in overdamped junctions, in contrast to the Josephson diode effect \cite{volkov_diode}.}

Overall, we have demonstrated that AC driving of the twisted interface can lead to sharp transitions between symmetry-broken and symmetric phases, distinguished by the presence or absence of second harmonic in the voltage. Similar phenomena have been recently predicted for Josephson loops and SQUIDs under microwave driving \cite{yerzhakov2024}; our results highlight the possibility of driving dynamical phase transitions using conventional microwave electronics.

\section{Implications for experiments on twisted  $\text{Bi$_2$Sr$_2$CaCu$_2$O$_{8+x}$}$}
\label{sec:exp}

We now discuss the application of our findings to experiments on twisted \bscco. Let us start with the Josephson plasma frequency. Away from the TRSB transition, but close to $\theta= 45^\circ$ the Josephson plasma frequency can be estimated as arising from purely second harmonic of the Josephson current, $\rm{Re} [\omega_{pl}^{(2)}\red{(T)}] = \frac{\sqrt{8\beta_c\red{(T)} - 1}}{(2\beta_c\red{(T)})} \frac{2 e J_{c2}\red{(T)} R}{\hbar}  $. While $ J_{c2} R \ll  J_{c1} (\theta=0^\circ) R$ \cite{pixley2025rev}, the interface junctions have been found to have considerably smaller $\beta_c\sim 1$ than the bulk (where values up to $10^4$ have been suggested \cite{kleiner1994,kleiner2000}). As a result, the estimated plasma frequency around $\theta=45^\circ$ would be $\gtrsim 0.1$ THz for current junctions (taking $J_{c2} R$ to be $\sim 0.2$ mV from 44.9$^\circ$ junctions in \cite{Zhao2023}), comparable to intrinsic zero-twist junctions \cite{plasmon_exp_1999,plasma2}. 

Such frequencies can be hard to measure with conventional RF electronics \cite{plasma2}. However, as discussed in Sec. \ref{sec:critlin}, the interface plasmon would soften to zero frequency at the TRSB transition. Even if explicit TRSB, leading to the "memory effect", is present, a considerable softening is expected that can be further turned into a complete one with a DC current bias, see Sec. \ref{subsec:memory}. One can roughly estimate the range of parameters where the plasmon will be at below, e.g., $10$ GHz. For temperature, using the Ginzburg-Landau temperature dependence of $J_{c1}, J_{c2}$, one expects a range of $\Delta T /T_c \approx10$ GHz/$0.1$ THz $= 0.1$, leading to $\Delta T \approx 8$ K for $T_c\approx 80$ K in \bscco. Furthermore, even if the peak in $Z_1$ occurs at too high a frequency, its behavior at low $\omega$ is also telling. Indeed, tuning towards the TRSB transition leads, Fig. \ref{fig:Z^1}, to an increase of low-frequency response, while going away from the transition point suppresses it, allowing to better locate the TRSB transition in experimental parameter space.

For the second-harmonic generation, one can estimate the maximal second harmonic voltage that can be generated before AC drive-induced transition is triggered. For the AC-drive induced phase transitions, the values of $I_0$ required are comparable to $J_{c2}$ and thus easily reachable in experiments, at least for the first dynamical phase transition from TRSB to a symmetric phase. Using the numerical solutions Fig. \ref{fig:max V2} (see also Appendix \ref{appendix:estimate} for details) we estimate the maximal $V_2$ to occur roughly in the middle of the TRSB region in twist angle and be of the order $0.3 J_{c2} R$, translating into $V_2 \geq 20 \mu$V at frequencies about $50$ GHz for $\beta_c \approx 2$. \pv{To obtain these voltage values we estimated $J_{c2} R$ taking the value of $RI_{c,30K}$ for $\theta$ in the range of $44.5^\circ-46.3^\circ$ in \cite{Zhao2023}; the frequency corresponds to the one where $V_2$ is found to be maximal in numerical solutions.} Decreasing the driving frequency by an order of magnitude, per Fig. \ref{fig:max V2} (a), would decrease $V_{2}$ also by an order of magnitude, potentially still allowing its observation with lock-in techniques.

\section{Outlook and Conclusions}
\label{sec:outlook}

In this work we have demonstrated that collective dynamics of twisted nodal superconductor interfaces bears critical information about the nature and properties of the time-reversal breaking phase.

First, we have shown that a soft Josephson plasmon emerges at the phase boundary of the TRSB phase (Fig. \ref{fig:phase-diagram1} and Sec. \ref{softmode}), which can reveal and quantify any non-spontaneous symmetry breaking (``memory effect" \cite{volkov_diode}), Sec. \ref{subsec:memory}, and can be detected using low-frequency AC impedance measurements (Sec. \ref{sec:linimped}). 

Secondly, we have demonstrated that nonlinear second-harmonic \pv{voltage} generation \pv{under AC current driving} due to the order parameter dynamics is a necessary and sufficient criterion for \red{time-reversal breaking, irrespective of the possible underlying microscopic mechanism,}  (Sec. \ref{sec: reentrance}) offering a convenient alternative to the diode effect for twisted superconductors. 

Finally, we have shown that nonlinear dynamics of the interface under strong AC driving can be described in terms of dynamical phase transitions between phases that can be sharply distinguished by the presence of a finite voltage response at the second harmonic of the drive frequency, Fig. \ref{fig:phase-diagram2} and Sec. \ref{sec:non-eq}.

Our estimates demonstrate that observation of these effects is within reach in twisted interfaces of \bscco, Sec. \ref{sec:exp}.

Our work establishes order parameter dynamics for twisted nodal superconductors as a means to both probe and control these systems. The phenomenological description developed here relies only on the presence of the Josephson effect with two harmonics, suggesting that our results may also find application to a wide variety of other platforms.

\section*{Acknowledgments}
We thank D. Kaplan, M. Franz, J. Pixley,  X.~Cui and P.~Kim for useful discussions and collaborations on related work. This work has been supported by by a Quantum-CT Quantum Regional Partnership Investments (QRPI) Award.

\appendix

\section{Twist angle for maximum second harmonic response}\label{appendix:estimate}

We estimate the maximal effect size of $\bar{V}_2$ numerically, as the maximum occurs outside of the high-frequency regime. Indeed, despite the increasing $I_0^{cr}$ for higher frequencies, we can see from the Jacobi-Anger expansion that second harmonic scales as $(\bar{\omega})^{-1/2}$. Thus, we do not expect this region to contain the maximal second harmonic voltage. 

\begin{figure}[h]
    \centering
    \includegraphics[width=1\linewidth]{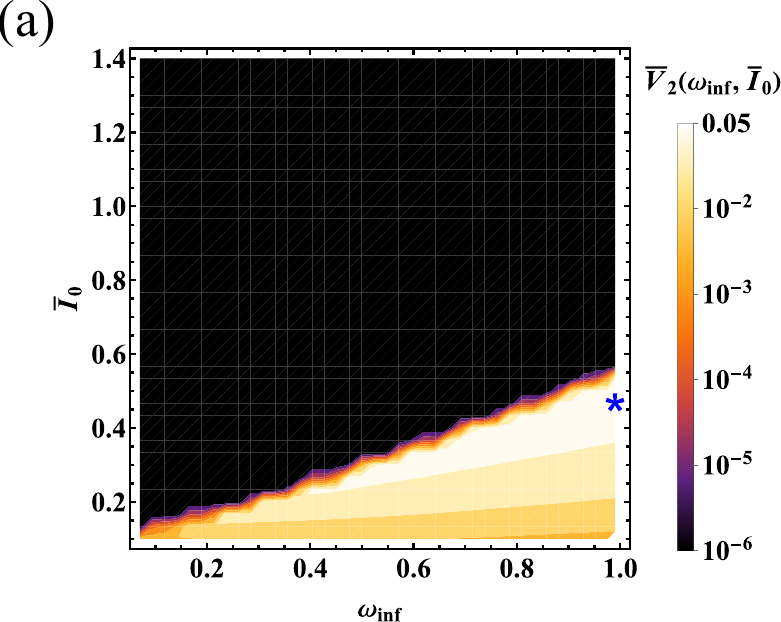}
    \includegraphics[width=1\linewidth]{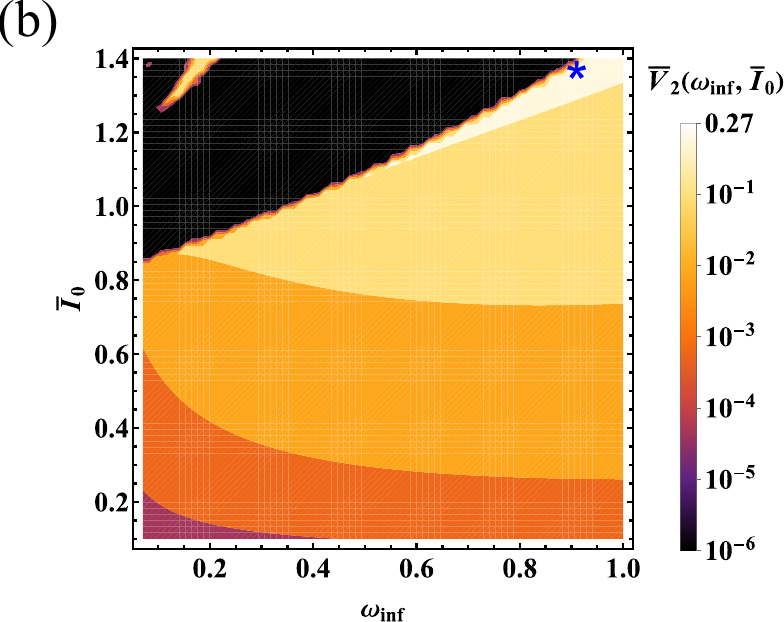}
    \caption{$\bar{V}_2(\bar{I_0},\omega_{\text{inf}})$ at $T=0$ for \red{$\bar{J}^{T=0}_{c1}=15$}, (a) $\theta=42^\circ$ and (b) $\theta=44.5^\circ$ \pv{(both within TRSB phase in equilibrium)} in the $\beta_c\rightarrow0$ limit. Note that $\omega=\omega_{\text{inf}}(2eJ_{c2}R/\hbar) $.The blue stars correspond to the maximum values in the range of frequencies and driving current displayed. The legend is bounded from above by the maximum value in the range.}
    \label{fig:appendix varying twist}
\end{figure}

The second harmonic response magnitude also depends on the twist angle. Since it vanishes both at $\theta=45^\circ$ and $\theta=\theta_c^\pm$, we took a value in between, $\theta=44^\circ$ (using estimated $\bar{J}_{c1} = 15$. Indeed, comparing the maximal $\bar{V}_2$ in the $\beta_c\to 0$ limit for $\theta=42^\circ$ and $\theta=44.5^\circ$, as shown in Fig. \ref{fig:appendix varying twist}, we see only a small increase in in the maximum response from 0.24 to 0.27 in the latter case.


We now discuss the $\beta_c$ dependence. For our numerical results, we have used $\bar{J}^{\red{T=0}}_{c1}=15$ comparable to the ratio between the critical current densities at $\theta=0^\circ$ and $\theta=45^\circ$ in Ref. \onlinecite{Zhao2023}, which varies between  $10-30$. 
\red{Hereafter, we take $T=0$.} For $\beta^{\red{T=0}}_c=2$, $\max[V_2]= (RJ_{c2})\cdot0.11$ occurs at $\bar{\omega} =0.85$ and $\bar{I}_0=0.65$.  For the $\beta^{\red{T=0}}_c=25$ case, we find $\max[V_2]=(RJ_{c2})\cdot0.05 $ occurring at $\bar{I}_0=0.55$ and $\bar{\omega}=0.55$.  And for $\beta\rightarrow 0$, we find $\max [V_2]=RJ_{c2}\cdot0.24$, with $\bar{\omega} =0.99$ and $\bar{I}_0=1.24$. We calculated the prefactors using the results in \cite{Zhao2023}, taking the value of $RI_{c,30K}$ for $\theta$ in the range of $44.5^\circ-46.3^\circ$. We find $RI_{C,30K}$ to be in the range of $0.2 \ \text{mV}- 2 \ \text{mV}$.

In real units, the frequency is given by $\bar{\omega}/(t_0\sqrt{\beta^{\red{T=0}}_c})$ and the current  $\bar{I}_0\cdot J_{c2}$. For $\beta_c\rightarrow 0$, the conversion to SI units is the same for the current, but is instead $\bar{\omega} / t_0$ for the frequency. The maximal second harmonic voltage that we expect to see in an experimental setting and its associated frequency and current values are given in Table \ref{tab:max V2}, they occur outside of the perturbative current regime and near the critical current, which can be seen in Fig. \ref{fig:max V2}.

\begin{table}[h]
    \centering
    \begin{tabular}{cccc}
        \hline
        $\beta_c$ & $V_2$(mV) & $\omega$(GHz) & $I_0$(mA) \\
        \hline
        0 & $0.05-0.5$& $86-918$ & $0.03-0.68$\\
        2   & $0.02-0.2$   & $52-556$   & $0.03-0.35$   \\
        25   & $0.01-0.1$   & $9.5-102$ & $0.02-0.3$\\
        \hline
    \end{tabular}
    \caption{The maximal second harmonic voltage for varying $\beta_c$ and its associated frequency and current.}
    \label{tab:max V2}
\end{table}

\section{Third harmonic response}
\label{app:3rd}

\begin{figure*}
        \subfigure{\includegraphics[width=0.325\linewidth]{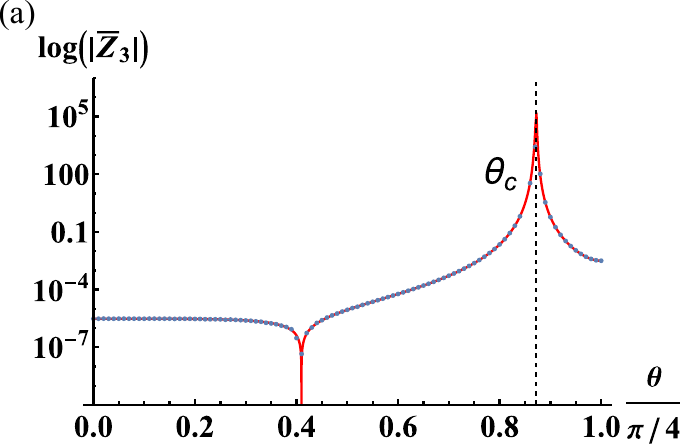}}
        \subfigure{\includegraphics[width=0.325\linewidth]{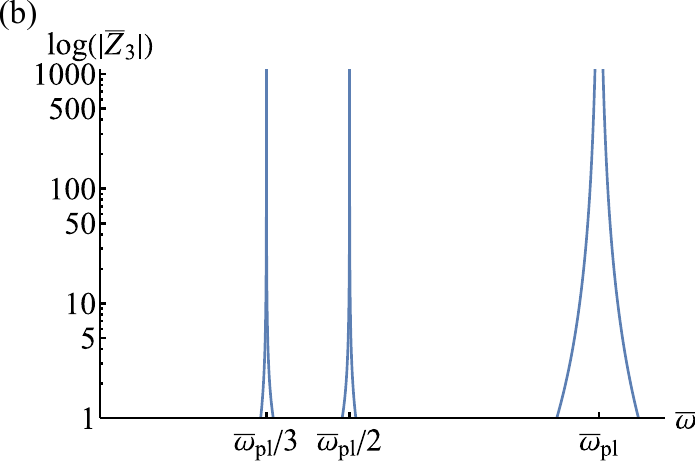}}
        \subfigure{\includegraphics[width=0.325\linewidth]{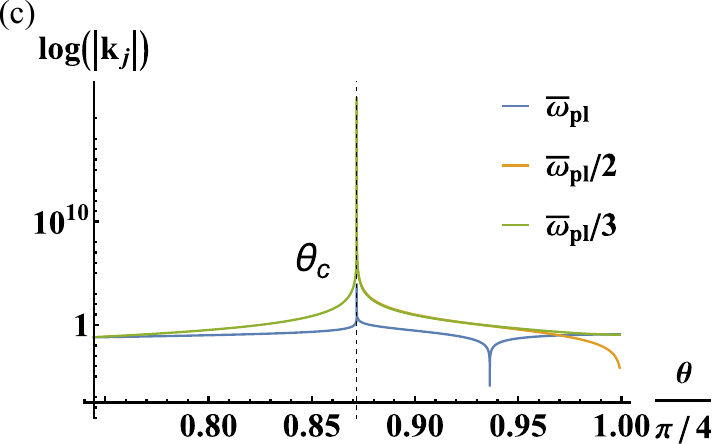}}
	\caption{Third harmonic impedance. (a) Absolute value of the third harmonic impedance in the small frequency limit. The red graph is the analytical result, Eq. \ref{3rd order impedance}. Blue points are obtained numerically for comparison. The vertical line corresponds to the critical angle for the onset of TRSB. The dip near $\theta=0.4\cdot(\pi/4)$ corresponds to the numerator of Eq. \ref{3rd order impedance} vanishing and does not correspond to a transition. (b) Log plot of the third order voltage in the $\beta_c\to 0$ limit in the TRSB phase for $\bar{J}^{\red{T=0}}_{c1}=10$, \red{$\theta=43^\circ$} and $T=0$. (c) Absolute values of the coefficients for three divergences of $Z_3$ as a function of $\theta$, Eq. \eqref{eq:kn}. 
	}
	\label{fig:third order impedance}
\end{figure*}

\red{We consider only the $T=0$ case for} the third order (in $I_0$) contribution to the phase difference. It contains the third harmonic part $\phi_3^{3\bar{\omega}} = \gamma_1e^{i3\bar{\omega} \tau} + \gamma_1^*e^{-i3\bar{\omega} \tau}$ where 
\begin{equation}
    \gamma_1 = \frac{-c(\theta)\alpha_1^3-2b(\theta)\beta_1\alpha_1}{f_3(\theta,\bar{\omega})}.
\end{equation}
The third harmonic impedance is then given by
\begin{equation}\label{3rd order impedance}
        |\bar{Z}_3(\theta,\bar{\omega})| = \left| \frac{3\bar{\omega}[2b^2(\theta)-c(\theta)f_2(\theta,\bar{\omega})]}{8f_1^3(\theta,\bar{\omega})f_2(\theta,\bar{\omega})f_3(\theta,\bar{\omega})}\right| 
\end{equation}
In the $\bar{\omega} \to 0$ limit (Fig. \ref{fig:third order impedance}), $|\bar{Z}_3|$ diverges at $\theta=\theta_c$ reflecting $\omega_{pl}\to 0$ at the transition. Furthermore, at a certain twist angle the numerator of Eq. \eqref{3rd order impedance} vanishes leading to zero third order response. This angle does not correspond to a transition, but rather demonstrates that third order response is nonuniversal and depends on the system's parameters.

Indeed, in the opposite limit $\bar{\omega} \gg \bar{\omega}_{pl}$, the third order voltage does not vanish for any twist angle, being present both in the symmetric and TRSB phase.

In the underdamped case i.e. $\beta_c \gg 1$, the third harmonic voltage will be strongly enhanced near various fractions of the plasma frequency, corresponding to values for which the denominator of Eq. \eqref{3rd order impedance} vanishes. In the TRSB phase and for when $\theta=45^\circ$, we find 3 different frequencies, $\bar{\omega}_{pl}$, $\bar{\omega}_{pl}/2$, and $\bar{\omega}_{pl} /3$ at which the voltage has a strong response. In the TRS phase, the system will only exhibit this behavior for $\bar{\omega}_{pl}$ and $\bar{\omega}_{pl} /3$, as $b(\theta)$ vanishes, leading to the denominator of $\bar{Z}_3$ containing only $f_1$ and $f_3$. This behavior is shown in Fig. \ref{fig:third order impedance} (b) for a system within TRSB phase. 

For $\beta_c\to 0$ we can further quantify the enhancement, by finding the coefficient of divergence:
\begin{equation}
k_n(\theta) = \bar{Z}_3(\bar{\omega})(\bar{\omega}_{pl}^2(\theta)-n^2\bar{\omega}^2)^{x_n}
\label{eq:kn}
\end{equation}
where $x_n = 3$ for $n=1$ and $x_j =1$ for $n=2,3$, as the divergences are of different orders. The results are shown in Fig. \ref{fig:third order impedance} (c). All divergences are enhanced at $\theta=\theta_c$, indicating enhanced resonant nonlinear response in the vicinity of the transition.


\section{Details of numerical calculations}
\label{sec:app_num}
To perform the Fourier transforms, we have integrated from $t=6000$ to at least $6000+250\cdot2\pi/\omega$ when the driving current extends into the nonperturbative regime. For perturbative calculations, we instead integrated from $t=100$ to at least $t=100+2\pi/\omega$ using $\bar{I}_0=0.01$.

\bibliography{ref}

\begin{thebibliography}{51}%
\makeatletter
\providecommand \@ifxundefined [1]{%
 \@ifx{#1\undefined}
}%
\providecommand \@ifnum [1]{%
 \ifnum #1\expandafter \@firstoftwo
 \else \expandafter \@secondoftwo
 \fi
}%
\providecommand \@ifx [1]{%
 \ifx #1\expandafter \@firstoftwo
 \else \expandafter \@secondoftwo
 \fi
}%
\providecommand \natexlab [1]{#1}%
\providecommand \enquote  [1]{``#1''}%
\providecommand \bibnamefont  [1]{#1}%
\providecommand \bibfnamefont [1]{#1}%
\providecommand \citenamefont [1]{#1}%
\providecommand \href@noop [0]{\@secondoftwo}%
\providecommand \href [0]{\begingroup \@sanitize@url \@href}%
\providecommand \@href[1]{\@@startlink{#1}\@@href}%
\providecommand \@@href[1]{\endgroup#1\@@endlink}%
\providecommand \@sanitize@url [0]{\catcode `\\12\catcode `\$12\catcode `\&12\catcode `\#12\catcode `\^12\catcode `\_12\catcode `\%12\relax}%
\providecommand \@@startlink[1]{}%
\providecommand \@@endlink[0]{}%
\providecommand \url  [0]{\begingroup\@sanitize@url \@url }%
\providecommand \@url [1]{\endgroup\@href {#1}{\urlprefix }}%
\providecommand \urlprefix  [0]{URL }%
\providecommand \Eprint [0]{\href }%
\providecommand \doibase [0]{https://doi.org/}%
\providecommand \selectlanguage [0]{\@gobble}%
\providecommand \bibinfo  [0]{\@secondoftwo}%
\providecommand \bibfield  [0]{\@secondoftwo}%
\providecommand \translation [1]{[#1]}%
\providecommand \BibitemOpen [0]{}%
\providecommand \bibitemStop [0]{}%
\providecommand \bibitemNoStop [0]{.\EOS\space}%
\providecommand \EOS [0]{\spacefactor3000\relax}%
\providecommand \BibitemShut  [1]{\csname bibitem#1\endcsname}%
\let\auto@bib@innerbib\@empty
\bibitem [{\citenamefont {Pixley}\ and\ \citenamefont {Volkov}(2025)}]{pixley2025rev}%
  \BibitemOpen
  \bibfield  {author} {\bibinfo {author} {\bibfnamefont {J.~H.}\ \bibnamefont {Pixley}}\ and\ \bibinfo {author} {\bibfnamefont {P.~A.}\ \bibnamefont {Volkov}},\ }\href {https://arxiv.org/abs/2503.23683} {\bibinfo {title} {Twisted nodal superconductors}} (\bibinfo {year} {2025}),\ \Eprint {https://arxiv.org/abs/2503.23683} {arXiv:2503.23683 [cond-mat.supr-con]} \BibitemShut {NoStop}%
\bibitem [{\citenamefont {Can}\ \emph {et~al.}(2021)\citenamefont {Can}, \citenamefont {Tummuru}, \citenamefont {Day}, \citenamefont {Elfimov}, \citenamefont {Damascelli},\ and\ \citenamefont {Franz}}]{Can2021}%
  \BibitemOpen
  \bibfield  {author} {\bibinfo {author} {\bibfnamefont {O.}~\bibnamefont {Can}}, \bibinfo {author} {\bibfnamefont {T.}~\bibnamefont {Tummuru}}, \bibinfo {author} {\bibfnamefont {R.~P.}\ \bibnamefont {Day}}, \bibinfo {author} {\bibfnamefont {I.}~\bibnamefont {Elfimov}}, \bibinfo {author} {\bibfnamefont {A.}~\bibnamefont {Damascelli}},\ and\ \bibinfo {author} {\bibfnamefont {M.}~\bibnamefont {Franz}},\ }\bibfield  {title} {\bibinfo {title} {High-temperature topological superconductivity in twisted double-layer copper oxides},\ }\href {https://doi.org/10.1038/s41567-020-01142-7} {\bibfield  {journal} {\bibinfo  {journal} {Nature Physics}\ }\textbf {\bibinfo {volume} {17}},\ \bibinfo {pages} {519} (\bibinfo {year} {2021})}\BibitemShut {NoStop}%
\bibitem [{\citenamefont {Volkov}\ \emph {et~al.}(2023)\citenamefont {Volkov}, \citenamefont {Wilson}, \citenamefont {Lucht},\ and\ \citenamefont {Pixley}}]{VolkovPRL-2023}%
  \BibitemOpen
  \bibfield  {author} {\bibinfo {author} {\bibfnamefont {P.~A.}\ \bibnamefont {Volkov}}, \bibinfo {author} {\bibfnamefont {J.~H.}\ \bibnamefont {Wilson}}, \bibinfo {author} {\bibfnamefont {K.~P.}\ \bibnamefont {Lucht}},\ and\ \bibinfo {author} {\bibfnamefont {J.~H.}\ \bibnamefont {Pixley}},\ }\bibfield  {title} {\bibinfo {title} {Current- and field-induced topology in twisted nodal superconductors},\ }\href {https://doi.org/10.1103/PhysRevLett.130.186001} {\bibfield  {journal} {\bibinfo  {journal} {Phys. Rev. Lett.}\ }\textbf {\bibinfo {volume} {130}},\ \bibinfo {pages} {186001} (\bibinfo {year} {2023})}\BibitemShut {NoStop}%
\bibitem [{\citenamefont {Tummuru}\ \emph {et~al.}(2021)\citenamefont {Tummuru}, \citenamefont {Can},\ and\ \citenamefont {Franz}}]{Tummuru-Franz-2021}%
  \BibitemOpen
  \bibfield  {author} {\bibinfo {author} {\bibfnamefont {T.}~\bibnamefont {Tummuru}}, \bibinfo {author} {\bibfnamefont {O.}~\bibnamefont {Can}},\ and\ \bibinfo {author} {\bibfnamefont {M.}~\bibnamefont {Franz}},\ }\bibfield  {title} {\bibinfo {title} {Chiral $p$-wave superconductivity in a twisted array of proximitized quantum wires},\ }\href {https://doi.org/10.1103/PhysRevB.103.L100501} {\bibfield  {journal} {\bibinfo  {journal} {Phys. Rev. B}\ }\textbf {\bibinfo {volume} {103}},\ \bibinfo {pages} {L100501} (\bibinfo {year} {2021})}\BibitemShut {NoStop}%
\bibitem [{\citenamefont {Volkov}\ \emph {et~al.}(2025)\citenamefont {Volkov}, \citenamefont {Zhao}, \citenamefont {Poccia}, \citenamefont {Cui}, \citenamefont {Kim},\ and\ \citenamefont {Pixley}}]{volkov_2025}%
  \BibitemOpen
  \bibfield  {author} {\bibinfo {author} {\bibfnamefont {P.~A.}\ \bibnamefont {Volkov}}, \bibinfo {author} {\bibfnamefont {S.~Y.~F.}\ \bibnamefont {Zhao}}, \bibinfo {author} {\bibfnamefont {N.}~\bibnamefont {Poccia}}, \bibinfo {author} {\bibfnamefont {X.}~\bibnamefont {Cui}}, \bibinfo {author} {\bibfnamefont {P.}~\bibnamefont {Kim}},\ and\ \bibinfo {author} {\bibfnamefont {J.~H.}\ \bibnamefont {Pixley}},\ }\bibfield  {title} {\bibinfo {title} {Josephson effects in twisted nodal superconductors},\ }\href {https://doi.org/10.1103/PhysRevB.111.014514} {\bibfield  {journal} {\bibinfo  {journal} {Phys. Rev. B}\ }\textbf {\bibinfo {volume} {111}},\ \bibinfo {pages} {014514} (\bibinfo {year} {2025})}\BibitemShut {NoStop}%
\bibitem [{\citenamefont {Volkov}\ \emph {et~al.}(2024)\citenamefont {Volkov}, \citenamefont {Lantagne-Hurtubise}, \citenamefont {Tummuru}, \citenamefont {Plugge}, \citenamefont {Pixley},\ and\ \citenamefont {Franz}}]{volkov_diode}%
  \BibitemOpen
  \bibfield  {author} {\bibinfo {author} {\bibfnamefont {P.~A.}\ \bibnamefont {Volkov}}, \bibinfo {author} {\bibfnamefont {E.}~\bibnamefont {Lantagne-Hurtubise}}, \bibinfo {author} {\bibfnamefont {T.}~\bibnamefont {Tummuru}}, \bibinfo {author} {\bibfnamefont {S.}~\bibnamefont {Plugge}}, \bibinfo {author} {\bibfnamefont {J.~H.}\ \bibnamefont {Pixley}},\ and\ \bibinfo {author} {\bibfnamefont {M.}~\bibnamefont {Franz}},\ }\bibfield  {title} {\bibinfo {title} {Josephson diode effects in twisted nodal superconductors},\ }\href {https://doi.org/10.1103/PhysRevB.109.094518} {\bibfield  {journal} {\bibinfo  {journal} {Phys. Rev. B}\ }\textbf {\bibinfo {volume} {109}},\ \bibinfo {pages} {094518} (\bibinfo {year} {2024})}\BibitemShut {NoStop}%
\bibitem [{\citenamefont {Zhao}\ \emph {et~al.}(2023)\citenamefont {Zhao}, \citenamefont {Cui}, \citenamefont {Volkov}, \citenamefont {Yoo}, \citenamefont {Lee}, \citenamefont {Gardener}, \citenamefont {Akey}, \citenamefont {Engelke}, \citenamefont {Ronen}, \citenamefont {Zhong}, \citenamefont {Gu}, \citenamefont {Plugge}, \citenamefont {Tummuru}, \citenamefont {Kim}, \citenamefont {Franz}, \citenamefont {Pixley}, \citenamefont {Poccia},\ and\ \citenamefont {Kim}}]{Zhao2023}%
  \BibitemOpen
  \bibfield  {author} {\bibinfo {author} {\bibfnamefont {S.~Y.~F.}\ \bibnamefont {Zhao}}, \bibinfo {author} {\bibfnamefont {X.}~\bibnamefont {Cui}}, \bibinfo {author} {\bibfnamefont {P.~A.}\ \bibnamefont {Volkov}}, \bibinfo {author} {\bibfnamefont {H.}~\bibnamefont {Yoo}}, \bibinfo {author} {\bibfnamefont {S.}~\bibnamefont {Lee}}, \bibinfo {author} {\bibfnamefont {J.~A.}\ \bibnamefont {Gardener}}, \bibinfo {author} {\bibfnamefont {A.~J.}\ \bibnamefont {Akey}}, \bibinfo {author} {\bibfnamefont {R.}~\bibnamefont {Engelke}}, \bibinfo {author} {\bibfnamefont {Y.}~\bibnamefont {Ronen}}, \bibinfo {author} {\bibfnamefont {R.}~\bibnamefont {Zhong}}, \bibinfo {author} {\bibfnamefont {G.}~\bibnamefont {Gu}}, \bibinfo {author} {\bibfnamefont {S.}~\bibnamefont {Plugge}}, \bibinfo {author} {\bibfnamefont {T.}~\bibnamefont {Tummuru}}, \bibinfo {author} {\bibfnamefont {M.}~\bibnamefont {Kim}}, \bibinfo {author} {\bibfnamefont {M.}~\bibnamefont {Franz}}, \bibinfo {author} {\bibfnamefont {J.~H.}\ \bibnamefont {Pixley}},
  \bibinfo {author} {\bibfnamefont {N.}~\bibnamefont {Poccia}},\ and\ \bibinfo {author} {\bibfnamefont {P.}~\bibnamefont {Kim}},\ }\bibfield  {title} {\bibinfo {title} {Time-reversal symmetry breaking superconductivity between twisted cuprate superconductors},\ }\href {https://doi.org/10.1126/science.abl8371} {\bibfield  {journal} {\bibinfo  {journal} {Science}\ }\textbf {\bibinfo {volume} {382}},\ \bibinfo {pages} {1422} (\bibinfo {year} {2023})},\ \Eprint {https://arxiv.org/abs/https://www.science.org/doi/pdf/10.1126/science.abl8371} {https://www.science.org/doi/pdf/10.1126/science.abl8371} \BibitemShut {NoStop}%
\bibitem [{\citenamefont {Zhu}\ \emph {et~al.}(2023)\citenamefont {Zhu}, \citenamefont {Wang}, \citenamefont {Wang}, \citenamefont {Hu}, \citenamefont {Gu}, \citenamefont {Zhu}, \citenamefont {Zhang},\ and\ \citenamefont {Xue}}]{xue_2023_OP}%
  \BibitemOpen
  \bibfield  {author} {\bibinfo {author} {\bibfnamefont {Y.}~\bibnamefont {Zhu}}, \bibinfo {author} {\bibfnamefont {H.}~\bibnamefont {Wang}}, \bibinfo {author} {\bibfnamefont {Z.}~\bibnamefont {Wang}}, \bibinfo {author} {\bibfnamefont {S.}~\bibnamefont {Hu}}, \bibinfo {author} {\bibfnamefont {G.}~\bibnamefont {Gu}}, \bibinfo {author} {\bibfnamefont {J.}~\bibnamefont {Zhu}}, \bibinfo {author} {\bibfnamefont {D.}~\bibnamefont {Zhang}},\ and\ \bibinfo {author} {\bibfnamefont {Q.-K.}\ \bibnamefont {Xue}},\ }\bibfield  {title} {\bibinfo {title} {Persistent josephson tunneling between ${\mathrm{bi}}_{2}{\mathrm{sr}}_{2}{\mathrm{cacu}}_{2}{\mathrm{o}}_{8+x}$ flakes twisted by ${45}^{\ensuremath{\circ}}$ across the superconducting dome},\ }\href {https://doi.org/10.1103/PhysRevB.108.174508} {\bibfield  {journal} {\bibinfo  {journal} {Phys. Rev. B}\ }\textbf {\bibinfo {volume} {108}},\ \bibinfo {pages} {174508} (\bibinfo {year} {2023})}\BibitemShut {NoStop}%
\bibitem [{\citenamefont {Qi}\ \emph {et~al.}(2025)\citenamefont {Qi}, \citenamefont {Ge}, \citenamefont {Ji}, \citenamefont {Ai}, \citenamefont {Ma}, \citenamefont {Wang}, \citenamefont {Cui}, \citenamefont {Liu}, \citenamefont {Wang},\ and\ \citenamefont {Wang}}]{diode_1flake_2025}%
  \BibitemOpen
  \bibfield  {author} {\bibinfo {author} {\bibfnamefont {S.}~\bibnamefont {Qi}}, \bibinfo {author} {\bibfnamefont {J.}~\bibnamefont {Ge}}, \bibinfo {author} {\bibfnamefont {C.}~\bibnamefont {Ji}}, \bibinfo {author} {\bibfnamefont {Y.}~\bibnamefont {Ai}}, \bibinfo {author} {\bibfnamefont {G.}~\bibnamefont {Ma}}, \bibinfo {author} {\bibfnamefont {Z.}~\bibnamefont {Wang}}, \bibinfo {author} {\bibfnamefont {Z.}~\bibnamefont {Cui}}, \bibinfo {author} {\bibfnamefont {Y.}~\bibnamefont {Liu}}, \bibinfo {author} {\bibfnamefont {Z.}~\bibnamefont {Wang}},\ and\ \bibinfo {author} {\bibfnamefont {J.}~\bibnamefont {Wang}},\ }\bibfield  {title} {\bibinfo {title} {High-temperature field-free superconducting diode effect in high-t c cuprates},\ }\href@noop {} {\bibfield  {journal} {\bibinfo  {journal} {Nature Communications}\ }\textbf {\bibinfo {volume} {16}},\ \bibinfo {pages} {531} (\bibinfo {year} {2025})}\BibitemShut {NoStop}%
\bibitem [{\citenamefont {Scott}(1974)}]{scott_1974}%
  \BibitemOpen
  \bibfield  {author} {\bibinfo {author} {\bibfnamefont {J.~F.}\ \bibnamefont {Scott}},\ }\bibfield  {title} {\bibinfo {title} {Soft-mode spectroscopy: Experimental studies of structural phase transitions},\ }\href {https://doi.org/10.1103/RevModPhys.46.83} {\bibfield  {journal} {\bibinfo  {journal} {Rev. Mod. Phys.}\ }\textbf {\bibinfo {volume} {46}},\ \bibinfo {pages} {83} (\bibinfo {year} {1974})}\BibitemShut {NoStop}%
\bibitem [{\citenamefont {Hohenberg}\ and\ \citenamefont {Halperin}(1977)}]{halperinhohenberg}%
  \BibitemOpen
  \bibfield  {author} {\bibinfo {author} {\bibfnamefont {P.~C.}\ \bibnamefont {Hohenberg}}\ and\ \bibinfo {author} {\bibfnamefont {B.~I.}\ \bibnamefont {Halperin}},\ }\bibfield  {title} {\bibinfo {title} {Theory of dynamic critical phenomena},\ }\href {https://doi.org/10.1103/RevModPhys.49.435} {\bibfield  {journal} {\bibinfo  {journal} {Rev. Mod. Phys.}\ }\textbf {\bibinfo {volume} {49}},\ \bibinfo {pages} {435} (\bibinfo {year} {1977})}\BibitemShut {NoStop}%
\bibitem [{\citenamefont {Kamba}(2021)}]{Kamba2021}%
  \BibitemOpen
  \bibfield  {author} {\bibinfo {author} {\bibfnamefont {S.}~\bibnamefont {Kamba}},\ }\bibfield  {title} {\bibinfo {title} {Soft-mode spectroscopy of ferroelectrics and multiferroics: A review},\ }\bibfield  {journal} {\bibinfo  {journal} {APL Materials}\ }\textbf {\bibinfo {volume} {9}},\ \href {https://doi.org/10.1063/5.0036066} {10.1063/5.0036066} (\bibinfo {year} {2021})\BibitemShut {NoStop}%
\bibitem [{\citenamefont {Volkov}\ \emph {et~al.}(2021{\natexlab{a}})\citenamefont {Volkov}, \citenamefont {Ye}, \citenamefont {Lohani}, \citenamefont {Feldman}, \citenamefont {Kanigel},\ and\ \citenamefont {Blumberg}}]{TNS}%
  \BibitemOpen
  \bibfield  {author} {\bibinfo {author} {\bibfnamefont {P.~A.}\ \bibnamefont {Volkov}}, \bibinfo {author} {\bibfnamefont {M.}~\bibnamefont {Ye}}, \bibinfo {author} {\bibfnamefont {H.}~\bibnamefont {Lohani}}, \bibinfo {author} {\bibfnamefont {I.}~\bibnamefont {Feldman}}, \bibinfo {author} {\bibfnamefont {A.}~\bibnamefont {Kanigel}},\ and\ \bibinfo {author} {\bibfnamefont {G.}~\bibnamefont {Blumberg}},\ }\bibfield  {title} {\bibinfo {title} {Critical charge fluctuations and emergent coherence in a strongly correlated excitonic insulator},\ }\href {https://doi.org/10.1038/s41535-021-00351-4} {\bibfield  {journal} {\bibinfo  {journal} {npj Quantum Materials}\ }\textbf {\bibinfo {volume} {6}},\ \bibinfo {pages} {52} (\bibinfo {year} {2021}{\natexlab{a}})}\BibitemShut {NoStop}%
\bibitem [{\citenamefont {Kim}\ \emph {et~al.}(2021)\citenamefont {Kim}, \citenamefont {Kim}, \citenamefont {Kim}, \citenamefont {Kwon}, \citenamefont {Kim},\ and\ \citenamefont {Kim}}]{tns_kim}%
  \BibitemOpen
  \bibfield  {author} {\bibinfo {author} {\bibfnamefont {K.}~\bibnamefont {Kim}}, \bibinfo {author} {\bibfnamefont {H.}~\bibnamefont {Kim}}, \bibinfo {author} {\bibfnamefont {J.}~\bibnamefont {Kim}}, \bibinfo {author} {\bibfnamefont {C.}~\bibnamefont {Kwon}}, \bibinfo {author} {\bibfnamefont {J.~S.}\ \bibnamefont {Kim}},\ and\ \bibinfo {author} {\bibfnamefont {B.~J.}\ \bibnamefont {Kim}},\ }\bibfield  {title} {\bibinfo {title} {Direct observation of excitonic instability in ta2nise5},\ }\bibfield  {journal} {\bibinfo  {journal} {Nature Communications}\ }\textbf {\bibinfo {volume} {12}},\ \href {https://doi.org/10.1038/s41467-021-22133-z} {10.1038/s41467-021-22133-z} (\bibinfo {year} {2021})\BibitemShut {NoStop}%
\bibitem [{\citenamefont {Ye}\ \emph {et~al.}(2021)\citenamefont {Ye}, \citenamefont {Volkov}, \citenamefont {Lohani}, \citenamefont {Feldman}, \citenamefont {Kim}, \citenamefont {Kanigel},\ and\ \citenamefont {Blumberg}}]{tnsmai}%
  \BibitemOpen
  \bibfield  {author} {\bibinfo {author} {\bibfnamefont {M.}~\bibnamefont {Ye}}, \bibinfo {author} {\bibfnamefont {P.~A.}\ \bibnamefont {Volkov}}, \bibinfo {author} {\bibfnamefont {H.}~\bibnamefont {Lohani}}, \bibinfo {author} {\bibfnamefont {I.}~\bibnamefont {Feldman}}, \bibinfo {author} {\bibfnamefont {M.}~\bibnamefont {Kim}}, \bibinfo {author} {\bibfnamefont {A.}~\bibnamefont {Kanigel}},\ and\ \bibinfo {author} {\bibfnamefont {G.}~\bibnamefont {Blumberg}},\ }\bibfield  {title} {\bibinfo {title} {{Lattice dynamics of the excitonic insulator ${\mathrm{Ta}}_{2}\mathrm{Ni}{({\mathrm{Se}}_{1\ensuremath{-}x}{\mathrm{S}}_{x})}_{5}$}},\ }\href {https://doi.org/10.1103/PhysRevB.104.045102} {\bibfield  {journal} {\bibinfo  {journal} {Phys. Rev. B}\ }\textbf {\bibinfo {volume} {104}},\ \bibinfo {pages} {045102} (\bibinfo {year} {2021})}\BibitemShut {NoStop}%
\bibitem [{\citenamefont {Volkov}\ \emph {et~al.}(2021{\natexlab{b}})\citenamefont {Volkov}, \citenamefont {Ye}, \citenamefont {Lohani}, \citenamefont {Feldman}, \citenamefont {Kanigel},\ and\ \citenamefont {Blumberg}}]{tnsx}%
  \BibitemOpen
  \bibfield  {author} {\bibinfo {author} {\bibfnamefont {P.~A.}\ \bibnamefont {Volkov}}, \bibinfo {author} {\bibfnamefont {M.}~\bibnamefont {Ye}}, \bibinfo {author} {\bibfnamefont {H.}~\bibnamefont {Lohani}}, \bibinfo {author} {\bibfnamefont {I.}~\bibnamefont {Feldman}}, \bibinfo {author} {\bibfnamefont {A.}~\bibnamefont {Kanigel}},\ and\ \bibinfo {author} {\bibfnamefont {G.}~\bibnamefont {Blumberg}},\ }\bibfield  {title} {\bibinfo {title} {{Failed excitonic quantum phase transition in ${\mathrm{Ta}}_{2}\mathrm{Ni}({\mathrm{Se}}_{1\ensuremath{-}x}{\mathrm{S}}_{x}{)}_{5}$}},\ }\href {https://doi.org/10.1103/PhysRevB.104.L241103} {\bibfield  {journal} {\bibinfo  {journal} {Phys. Rev. B}\ }\textbf {\bibinfo {volume} {104}},\ \bibinfo {pages} {L241103} (\bibinfo {year} {2021}{\natexlab{b}})}\BibitemShut {NoStop}%
\bibitem [{\citenamefont {Kogar}\ \emph {et~al.}(2017)\citenamefont {Kogar}, \citenamefont {Rak}, \citenamefont {Vig}, \citenamefont {Husain}, \citenamefont {Flicker}, \citenamefont {Joe}, \citenamefont {Venema}, \citenamefont {MacDougall}, \citenamefont {Chiang}, \citenamefont {Fradkin}, \citenamefont {van Wezel},\ and\ \citenamefont {Abbamonte}}]{Kogar2017}%
  \BibitemOpen
  \bibfield  {author} {\bibinfo {author} {\bibfnamefont {A.}~\bibnamefont {Kogar}}, \bibinfo {author} {\bibfnamefont {M.~S.}\ \bibnamefont {Rak}}, \bibinfo {author} {\bibfnamefont {S.}~\bibnamefont {Vig}}, \bibinfo {author} {\bibfnamefont {A.~A.}\ \bibnamefont {Husain}}, \bibinfo {author} {\bibfnamefont {F.}~\bibnamefont {Flicker}}, \bibinfo {author} {\bibfnamefont {Y.~I.}\ \bibnamefont {Joe}}, \bibinfo {author} {\bibfnamefont {L.}~\bibnamefont {Venema}}, \bibinfo {author} {\bibfnamefont {G.~J.}\ \bibnamefont {MacDougall}}, \bibinfo {author} {\bibfnamefont {T.~C.}\ \bibnamefont {Chiang}}, \bibinfo {author} {\bibfnamefont {E.}~\bibnamefont {Fradkin}}, \bibinfo {author} {\bibfnamefont {J.}~\bibnamefont {van Wezel}},\ and\ \bibinfo {author} {\bibfnamefont {P.}~\bibnamefont {Abbamonte}},\ }\bibfield  {title} {\bibinfo {title} {Signatures of exciton condensation in a transition metal dichalcogenide},\ }\href {https://doi.org/10.1126/science.aam6432} {\bibfield  {journal} {\bibinfo  {journal} {Science}\ }\textbf
  {\bibinfo {volume} {358}},\ \bibinfo {pages} {1314–1317} (\bibinfo {year} {2017})}\BibitemShut {NoStop}%
\bibitem [{\citenamefont {Giamarchi}\ \emph {et~al.}(2008)\citenamefont {Giamarchi}, \citenamefont {R\"{u}egg},\ and\ \citenamefont {Tchernyshyov}}]{Giamarchi2008}%
  \BibitemOpen
  \bibfield  {author} {\bibinfo {author} {\bibfnamefont {T.}~\bibnamefont {Giamarchi}}, \bibinfo {author} {\bibfnamefont {C.}~\bibnamefont {R\"{u}egg}},\ and\ \bibinfo {author} {\bibfnamefont {O.}~\bibnamefont {Tchernyshyov}},\ }\bibfield  {title} {\bibinfo {title} {Bose–einstein condensation in magnetic insulators},\ }\href {https://doi.org/10.1038/nphys893} {\bibfield  {journal} {\bibinfo  {journal} {Nature Physics}\ }\textbf {\bibinfo {volume} {4}},\ \bibinfo {pages} {198–204} (\bibinfo {year} {2008})}\BibitemShut {NoStop}%
\bibitem [{\citenamefont {Lin}\ and\ \citenamefont {Hu}(2012)}]{lin_2012}%
  \BibitemOpen
  \bibfield  {author} {\bibinfo {author} {\bibfnamefont {S.-Z.}\ \bibnamefont {Lin}}\ and\ \bibinfo {author} {\bibfnamefont {X.}~\bibnamefont {Hu}},\ }\bibfield  {title} {\bibinfo {title} {Massless leggett mode in three-band superconductors with time-reversal-symmetry breaking},\ }\href {https://doi.org/10.1103/PhysRevLett.108.177005} {\bibfield  {journal} {\bibinfo  {journal} {Phys. Rev. Lett.}\ }\textbf {\bibinfo {volume} {108}},\ \bibinfo {pages} {177005} (\bibinfo {year} {2012})}\BibitemShut {NoStop}%
\bibitem [{\citenamefont {Stanev}(2012)}]{stanev_2012}%
  \BibitemOpen
  \bibfield  {author} {\bibinfo {author} {\bibfnamefont {V.}~\bibnamefont {Stanev}},\ }\bibfield  {title} {\bibinfo {title} {Model of collective modes in three-band superconductors with repulsive interband interactions},\ }\href {https://doi.org/10.1103/PhysRevB.85.174520} {\bibfield  {journal} {\bibinfo  {journal} {Phys. Rev. B}\ }\textbf {\bibinfo {volume} {85}},\ \bibinfo {pages} {174520} (\bibinfo {year} {2012})}\BibitemShut {NoStop}%
\bibitem [{\citenamefont {Carlstr\"om}\ \emph {et~al.}(2011)\citenamefont {Carlstr\"om}, \citenamefont {Garaud},\ and\ \citenamefont {Babaev}}]{babaev_2011}%
  \BibitemOpen
  \bibfield  {author} {\bibinfo {author} {\bibfnamefont {J.}~\bibnamefont {Carlstr\"om}}, \bibinfo {author} {\bibfnamefont {J.}~\bibnamefont {Garaud}},\ and\ \bibinfo {author} {\bibfnamefont {E.}~\bibnamefont {Babaev}},\ }\bibfield  {title} {\bibinfo {title} {Length scales, collective modes, and type-1.5 regimes in three-band superconductors},\ }\href {https://doi.org/10.1103/PhysRevB.84.134518} {\bibfield  {journal} {\bibinfo  {journal} {Phys. Rev. B}\ }\textbf {\bibinfo {volume} {84}},\ \bibinfo {pages} {134518} (\bibinfo {year} {2011})}\BibitemShut {NoStop}%
\bibitem [{\citenamefont {Maiti}\ and\ \citenamefont {Chubukov}(2013)}]{maiti2013}%
  \BibitemOpen
  \bibfield  {author} {\bibinfo {author} {\bibfnamefont {S.}~\bibnamefont {Maiti}}\ and\ \bibinfo {author} {\bibfnamefont {A.~V.}\ \bibnamefont {Chubukov}},\ }\bibfield  {title} {\bibinfo {title} {$s+is$ state with broken time-reversal symmetry in fe-based superconductors},\ }\href {https://doi.org/10.1103/PhysRevB.87.144511} {\bibfield  {journal} {\bibinfo  {journal} {Phys. Rev. B}\ }\textbf {\bibinfo {volume} {87}},\ \bibinfo {pages} {144511} (\bibinfo {year} {2013})}\BibitemShut {NoStop}%
\bibitem [{\citenamefont {Marciani}\ \emph {et~al.}(2013)\citenamefont {Marciani}, \citenamefont {Fanfarillo}, \citenamefont {Castellani},\ and\ \citenamefont {Benfatto}}]{marciani2013}%
  \BibitemOpen
  \bibfield  {author} {\bibinfo {author} {\bibfnamefont {M.}~\bibnamefont {Marciani}}, \bibinfo {author} {\bibfnamefont {L.}~\bibnamefont {Fanfarillo}}, \bibinfo {author} {\bibfnamefont {C.}~\bibnamefont {Castellani}},\ and\ \bibinfo {author} {\bibfnamefont {L.}~\bibnamefont {Benfatto}},\ }\bibfield  {title} {\bibinfo {title} {Leggett modes in iron-based superconductors as a probe of time-reversal symmetry breaking},\ }\href {https://doi.org/10.1103/PhysRevB.88.214508} {\bibfield  {journal} {\bibinfo  {journal} {Phys. Rev. B}\ }\textbf {\bibinfo {volume} {88}},\ \bibinfo {pages} {214508} (\bibinfo {year} {2013})}\BibitemShut {NoStop}%
\bibitem [{\citenamefont {Zhao}\ \emph {et~al.}(2020)\citenamefont {Zhao}, \citenamefont {Song}, \citenamefont {Hu}, \citenamefont {Xie}, \citenamefont {Liu}, \citenamefont {Luo}, \citenamefont {Jiang}, \citenamefont {Zhang}, \citenamefont {Nie}, \citenamefont {Meng}, \citenamefont {Duan}, \citenamefont {Liu}, \citenamefont {Xie},\ and\ \citenamefont {Liu}}]{trsb_fe}%
  \BibitemOpen
  \bibfield  {author} {\bibinfo {author} {\bibfnamefont {S.~Z.}\ \bibnamefont {Zhao}}, \bibinfo {author} {\bibfnamefont {H.-Y.}\ \bibnamefont {Song}}, \bibinfo {author} {\bibfnamefont {L.~L.}\ \bibnamefont {Hu}}, \bibinfo {author} {\bibfnamefont {T.}~\bibnamefont {Xie}}, \bibinfo {author} {\bibfnamefont {C.}~\bibnamefont {Liu}}, \bibinfo {author} {\bibfnamefont {H.~Q.}\ \bibnamefont {Luo}}, \bibinfo {author} {\bibfnamefont {C.-Y.}\ \bibnamefont {Jiang}}, \bibinfo {author} {\bibfnamefont {X.}~\bibnamefont {Zhang}}, \bibinfo {author} {\bibfnamefont {X.~C.}\ \bibnamefont {Nie}}, \bibinfo {author} {\bibfnamefont {J.-Q.}\ \bibnamefont {Meng}}, \bibinfo {author} {\bibfnamefont {Y.-X.}\ \bibnamefont {Duan}}, \bibinfo {author} {\bibfnamefont {S.-B.}\ \bibnamefont {Liu}}, \bibinfo {author} {\bibfnamefont {H.-Y.}\ \bibnamefont {Xie}},\ and\ \bibinfo {author} {\bibfnamefont {H.~Y.}\ \bibnamefont {Liu}},\ }\bibfield  {title} {\bibinfo {title} {Observation of soft leggett mode in superconducting
  ${\mathrm{cakfe}}_{4}{\mathrm{as}}_{4}$},\ }\href {https://doi.org/10.1103/PhysRevB.102.144519} {\bibfield  {journal} {\bibinfo  {journal} {Phys. Rev. B}\ }\textbf {\bibinfo {volume} {102}},\ \bibinfo {pages} {144519} (\bibinfo {year} {2020})}\BibitemShut {NoStop}%
\bibitem [{\citenamefont {Shimano}\ and\ \citenamefont {Tsuji}(2020)}]{shimano2020}%
  \BibitemOpen
  \bibfield  {author} {\bibinfo {author} {\bibfnamefont {R.}~\bibnamefont {Shimano}}\ and\ \bibinfo {author} {\bibfnamefont {N.}~\bibnamefont {Tsuji}},\ }\bibfield  {title} {\bibinfo {title} {Higgs mode in superconductors},\ }\href {https://doi.org/https://doi.org/10.1146/annurev-conmatphys-031119-050813} {\bibfield  {journal} {\bibinfo  {journal} {Annual Review of Condensed Matter Physics}\ }\textbf {\bibinfo {volume} {11}},\ \bibinfo {pages} {103} (\bibinfo {year} {2020})}\BibitemShut {NoStop}%
\bibitem [{\citenamefont {Gabriele}\ \emph {et~al.}(2021)\citenamefont {Gabriele}, \citenamefont {Udina},\ and\ \citenamefont {Benfatto}}]{Gabriele2021}%
  \BibitemOpen
  \bibfield  {author} {\bibinfo {author} {\bibfnamefont {F.}~\bibnamefont {Gabriele}}, \bibinfo {author} {\bibfnamefont {M.}~\bibnamefont {Udina}},\ and\ \bibinfo {author} {\bibfnamefont {L.}~\bibnamefont {Benfatto}},\ }\bibfield  {title} {\bibinfo {title} {Non-linear terahertz driving of plasma waves in layered cuprates},\ }\bibfield  {journal} {\bibinfo  {journal} {Nature Communications}\ }\textbf {\bibinfo {volume} {12}},\ \href {https://doi.org/10.1038/s41467-021-21041-6} {10.1038/s41467-021-21041-6} (\bibinfo {year} {2021})\BibitemShut {NoStop}%
\bibitem [{\citenamefont {Fiore}\ \emph {et~al.}(2024)\citenamefont {Fiore}, \citenamefont {Sellati}, \citenamefont {Gabriele}, \citenamefont {Castellani}, \citenamefont {Seibold}, \citenamefont {Udina},\ and\ \citenamefont {Benfatto}}]{PhysRevB.110.L060504}%
  \BibitemOpen
  \bibfield  {author} {\bibinfo {author} {\bibfnamefont {J.}~\bibnamefont {Fiore}}, \bibinfo {author} {\bibfnamefont {N.}~\bibnamefont {Sellati}}, \bibinfo {author} {\bibfnamefont {F.}~\bibnamefont {Gabriele}}, \bibinfo {author} {\bibfnamefont {C.}~\bibnamefont {Castellani}}, \bibinfo {author} {\bibfnamefont {G.}~\bibnamefont {Seibold}}, \bibinfo {author} {\bibfnamefont {M.}~\bibnamefont {Udina}},\ and\ \bibinfo {author} {\bibfnamefont {L.}~\bibnamefont {Benfatto}},\ }\bibfield  {title} {\bibinfo {title} {Investigating josephson plasmons in layered cuprates via nonlinear terahertz spectroscopy},\ }\href {https://doi.org/10.1103/PhysRevB.110.L060504} {\bibfield  {journal} {\bibinfo  {journal} {Phys. Rev. B}\ }\textbf {\bibinfo {volume} {110}},\ \bibinfo {pages} {L060504} (\bibinfo {year} {2024})}\BibitemShut {NoStop}%
\bibitem [{\citenamefont {Katsumi}\ \emph {et~al.}(2023)\citenamefont {Katsumi}, \citenamefont {Nishida}, \citenamefont {Kaiser}, \citenamefont {Miyasaka}, \citenamefont {Tajima},\ and\ \citenamefont {Shimano}}]{Katsumi2023}%
  \BibitemOpen
  \bibfield  {author} {\bibinfo {author} {\bibfnamefont {K.}~\bibnamefont {Katsumi}}, \bibinfo {author} {\bibfnamefont {M.}~\bibnamefont {Nishida}}, \bibinfo {author} {\bibfnamefont {S.}~\bibnamefont {Kaiser}}, \bibinfo {author} {\bibfnamefont {S.}~\bibnamefont {Miyasaka}}, \bibinfo {author} {\bibfnamefont {S.}~\bibnamefont {Tajima}},\ and\ \bibinfo {author} {\bibfnamefont {R.}~\bibnamefont {Shimano}},\ }\bibfield  {title} {\bibinfo {title} {Near-infrared light-induced superconducting-like state in underdoped $\mathrm{Y}{\mathrm{ba}}_{2}{\mathrm{cu}}_{3}{\mathrm{o}}_{y}$ studied by $c$-axis terahertz third-harmonic generation},\ }\href {https://doi.org/10.1103/PhysRevB.107.214506} {\bibfield  {journal} {\bibinfo  {journal} {Phys. Rev. B}\ }\textbf {\bibinfo {volume} {107}},\ \bibinfo {pages} {214506} (\bibinfo {year} {2023})}\BibitemShut {NoStop}%
\bibitem [{\citenamefont {Rajasekaran}\ \emph {et~al.}(2018)\citenamefont {Rajasekaran}, \citenamefont {Okamoto}, \citenamefont {Mathey}, \citenamefont {Fechner}, \citenamefont {Thampy}, \citenamefont {Gu},\ and\ \citenamefont {Cavalleri}}]{Rajasekaran2018}%
  \BibitemOpen
  \bibfield  {author} {\bibinfo {author} {\bibfnamefont {S.}~\bibnamefont {Rajasekaran}}, \bibinfo {author} {\bibfnamefont {J.}~\bibnamefont {Okamoto}}, \bibinfo {author} {\bibfnamefont {L.}~\bibnamefont {Mathey}}, \bibinfo {author} {\bibfnamefont {M.}~\bibnamefont {Fechner}}, \bibinfo {author} {\bibfnamefont {V.}~\bibnamefont {Thampy}}, \bibinfo {author} {\bibfnamefont {G.~D.}\ \bibnamefont {Gu}},\ and\ \bibinfo {author} {\bibfnamefont {A.}~\bibnamefont {Cavalleri}},\ }\bibfield  {title} {\bibinfo {title} {Probing optically silent superfluid stripes in cuprates},\ }\href {https://doi.org/10.1126/science.aan3438} {\bibfield  {journal} {\bibinfo  {journal} {Science}\ }\textbf {\bibinfo {volume} {359}},\ \bibinfo {pages} {575} (\bibinfo {year} {2018})},\ \Eprint {https://arxiv.org/abs/https://www.science.org/doi/pdf/10.1126/science.aan3438} {https://www.science.org/doi/pdf/10.1126/science.aan3438} \BibitemShut {NoStop}%
\bibitem [{\citenamefont {G\'omez~Salvador}\ \emph {et~al.}(2024)\citenamefont {G\'omez~Salvador}, \citenamefont {Dolgirev}, \citenamefont {Michael}, \citenamefont {Liu}, \citenamefont {Pavicevic}, \citenamefont {Fechner}, \citenamefont {Cavalleri},\ and\ \citenamefont {Demler}}]{Salvador2024}%
  \BibitemOpen
  \bibfield  {author} {\bibinfo {author} {\bibfnamefont {A.}~\bibnamefont {G\'omez~Salvador}}, \bibinfo {author} {\bibfnamefont {P.~E.}\ \bibnamefont {Dolgirev}}, \bibinfo {author} {\bibfnamefont {M.~H.}\ \bibnamefont {Michael}}, \bibinfo {author} {\bibfnamefont {A.}~\bibnamefont {Liu}}, \bibinfo {author} {\bibfnamefont {D.}~\bibnamefont {Pavicevic}}, \bibinfo {author} {\bibfnamefont {M.}~\bibnamefont {Fechner}}, \bibinfo {author} {\bibfnamefont {A.}~\bibnamefont {Cavalleri}},\ and\ \bibinfo {author} {\bibfnamefont {E.}~\bibnamefont {Demler}},\ }\bibfield  {title} {\bibinfo {title} {Principles of two-dimensional terahertz spectroscopy of collective excitations: The case of josephson plasmons in layered superconductors},\ }\href {https://doi.org/10.1103/PhysRevB.110.094514} {\bibfield  {journal} {\bibinfo  {journal} {Phys. Rev. B}\ }\textbf {\bibinfo {volume} {110}},\ \bibinfo {pages} {094514} (\bibinfo {year} {2024})}\BibitemShut {NoStop}%
\bibitem [{\citenamefont {Kaj}\ \emph {et~al.}(2023)\citenamefont {Kaj}, \citenamefont {Cremin}, \citenamefont {Hammock}, \citenamefont {Schalch}, \citenamefont {Basov},\ and\ \citenamefont {Averitt}}]{Kaj2023}%
  \BibitemOpen
  \bibfield  {author} {\bibinfo {author} {\bibfnamefont {K.}~\bibnamefont {Kaj}}, \bibinfo {author} {\bibfnamefont {K.~A.}\ \bibnamefont {Cremin}}, \bibinfo {author} {\bibfnamefont {I.}~\bibnamefont {Hammock}}, \bibinfo {author} {\bibfnamefont {J.}~\bibnamefont {Schalch}}, \bibinfo {author} {\bibfnamefont {D.~N.}\ \bibnamefont {Basov}},\ and\ \bibinfo {author} {\bibfnamefont {R.~D.}\ \bibnamefont {Averitt}},\ }\bibfield  {title} {\bibinfo {title} {Terahertz third harmonic generation in $c$-axis ${\mathrm{la}}_{1.85}{\mathrm{sr}}_{0.15}{\mathrm{cuo}}_{4}$},\ }\href {https://doi.org/10.1103/PhysRevB.107.L140504} {\bibfield  {journal} {\bibinfo  {journal} {Phys. Rev. B}\ }\textbf {\bibinfo {volume} {107}},\ \bibinfo {pages} {L140504} (\bibinfo {year} {2023})}\BibitemShut {NoStop}%
\bibitem [{\citenamefont {Seibold}\ \emph {et~al.}(2021)\citenamefont {Seibold}, \citenamefont {Udina}, \citenamefont {Castellani},\ and\ \citenamefont {Benfatto}}]{Seibold2021}%
  \BibitemOpen
  \bibfield  {author} {\bibinfo {author} {\bibfnamefont {G.}~\bibnamefont {Seibold}}, \bibinfo {author} {\bibfnamefont {M.}~\bibnamefont {Udina}}, \bibinfo {author} {\bibfnamefont {C.}~\bibnamefont {Castellani}},\ and\ \bibinfo {author} {\bibfnamefont {L.}~\bibnamefont {Benfatto}},\ }\bibfield  {title} {\bibinfo {title} {Third harmonic generation from collective modes in disordered superconductors},\ }\href {https://doi.org/10.1103/PhysRevB.103.014512} {\bibfield  {journal} {\bibinfo  {journal} {Phys. Rev. B}\ }\textbf {\bibinfo {volume} {103}},\ \bibinfo {pages} {014512} (\bibinfo {year} {2021})}\BibitemShut {NoStop}%
\bibitem [{\citenamefont {Zhang}\ \emph {et~al.}(2023)\citenamefont {Zhang}, \citenamefont {Sun}, \citenamefont {Liu}, \citenamefont {Wang}, \citenamefont {Wu}, \citenamefont {Yue}, \citenamefont {Xu}, \citenamefont {Hu}, \citenamefont {Li}, \citenamefont {Zhou}, \citenamefont {Yuan}, \citenamefont {Gu}, \citenamefont {Dong},\ and\ \citenamefont {Wang}}]{Zhang2023NSR}%
  \BibitemOpen
  \bibfield  {author} {\bibinfo {author} {\bibfnamefont {S.}~\bibnamefont {Zhang}}, \bibinfo {author} {\bibfnamefont {Z.}~\bibnamefont {Sun}}, \bibinfo {author} {\bibfnamefont {Q.}~\bibnamefont {Liu}}, \bibinfo {author} {\bibfnamefont {Z.}~\bibnamefont {Wang}}, \bibinfo {author} {\bibfnamefont {Q.}~\bibnamefont {Wu}}, \bibinfo {author} {\bibfnamefont {L.}~\bibnamefont {Yue}}, \bibinfo {author} {\bibfnamefont {S.}~\bibnamefont {Xu}}, \bibinfo {author} {\bibfnamefont {T.}~\bibnamefont {Hu}}, \bibinfo {author} {\bibfnamefont {R.}~\bibnamefont {Li}}, \bibinfo {author} {\bibfnamefont {X.}~\bibnamefont {Zhou}}, \bibinfo {author} {\bibfnamefont {J.}~\bibnamefont {Yuan}}, \bibinfo {author} {\bibfnamefont {G.}~\bibnamefont {Gu}}, \bibinfo {author} {\bibfnamefont {T.}~\bibnamefont {Dong}},\ and\ \bibinfo {author} {\bibfnamefont {N.}~\bibnamefont {Wang}},\ }\bibfield  {title} {\bibinfo {title} {Revealing the frequency-dependent oscillations in the nonlinear terahertz response induced by the josephson current},\ }\href
  {https://doi.org/10.1093/nsr/nwad163} {\bibfield  {journal} {\bibinfo  {journal} {National Science Review}\ }\textbf {\bibinfo {volume} {10}},\ \bibinfo {pages} {nwad163} (\bibinfo {year} {2023})},\ \Eprint {https://arxiv.org/abs/https://academic.oup.com/nsr/article-pdf/10/11/nwad163/51961009/nwad163.pdf} {https://academic.oup.com/nsr/article-pdf/10/11/nwad163/51961009/nwad163.pdf} \BibitemShut {NoStop}%
\bibitem [{\citenamefont {Raj}\ \emph {et~al.}(2024)\citenamefont {Raj}, \citenamefont {Postlewaite}, \citenamefont {Chaudhary},\ and\ \citenamefont {Fiete}}]{fiete2024}%
  \BibitemOpen
  \bibfield  {author} {\bibinfo {author} {\bibfnamefont {A.}~\bibnamefont {Raj}}, \bibinfo {author} {\bibfnamefont {A.}~\bibnamefont {Postlewaite}}, \bibinfo {author} {\bibfnamefont {S.}~\bibnamefont {Chaudhary}},\ and\ \bibinfo {author} {\bibfnamefont {G.~A.}\ \bibnamefont {Fiete}},\ }\bibfield  {title} {\bibinfo {title} {Nonlinear optical responses in multiorbital topological superconductors},\ }\href {https://doi.org/10.1103/PhysRevB.109.184514} {\bibfield  {journal} {\bibinfo  {journal} {Phys. Rev. B}\ }\textbf {\bibinfo {volume} {109}},\ \bibinfo {pages} {184514} (\bibinfo {year} {2024})}\BibitemShut {NoStop}%
\bibitem [{\citenamefont {Kaplan}\ \emph {et~al.}(2025)\citenamefont {Kaplan}, \citenamefont {Lucht}, \citenamefont {Volkov},\ and\ \citenamefont {Pixley}}]{kaplan2025}%
  \BibitemOpen
  \bibfield  {author} {\bibinfo {author} {\bibfnamefont {D.}~\bibnamefont {Kaplan}}, \bibinfo {author} {\bibfnamefont {K.~P.}\ \bibnamefont {Lucht}}, \bibinfo {author} {\bibfnamefont {P.~A.}\ \bibnamefont {Volkov}},\ and\ \bibinfo {author} {\bibfnamefont {J.~H.}\ \bibnamefont {Pixley}},\ }\href {https://arxiv.org/abs/2502.12265} {\bibinfo {title} {Quantum geometric photocurrents of quasiparticles in superconductors}} (\bibinfo {year} {2025}),\ \Eprint {https://arxiv.org/abs/2502.12265} {arXiv:2502.12265 [cond-mat.supr-con]} \BibitemShut {NoStop}%
\bibitem [{\citenamefont {Yuan}\ \emph {et~al.}(2023)\citenamefont {Yuan}, \citenamefont {Vituri}, \citenamefont {Berg}, \citenamefont {Spivak},\ and\ \citenamefont {Kivelson}}]{yuan2023}%
  \BibitemOpen
  \bibfield  {author} {\bibinfo {author} {\bibfnamefont {A.~C.}\ \bibnamefont {Yuan}}, \bibinfo {author} {\bibfnamefont {Y.}~\bibnamefont {Vituri}}, \bibinfo {author} {\bibfnamefont {E.}~\bibnamefont {Berg}}, \bibinfo {author} {\bibfnamefont {B.}~\bibnamefont {Spivak}},\ and\ \bibinfo {author} {\bibfnamefont {S.~A.}\ \bibnamefont {Kivelson}},\ }\bibfield  {title} {\bibinfo {title} {Inhomogeneity-induced time-reversal symmetry breaking in cuprate twist junctions},\ }\href {https://doi.org/10.1103/PhysRevB.108.L100505} {\bibfield  {journal} {\bibinfo  {journal} {Phys. Rev. B}\ }\textbf {\bibinfo {volume} {108}},\ \bibinfo {pages} {L100505} (\bibinfo {year} {2023})}\BibitemShut {NoStop}%
\bibitem [{\citenamefont {Yuan}\ and\ \citenamefont {Crawford}(2024)}]{yuan2024}%
  \BibitemOpen
  \bibfield  {author} {\bibinfo {author} {\bibfnamefont {A.~C.}\ \bibnamefont {Yuan}}\ and\ \bibinfo {author} {\bibfnamefont {N.}~\bibnamefont {Crawford}},\ }\href {https://arxiv.org/abs/2412.17905} {\bibinfo {title} {Infinitely stable disordered systems on emergent fractal structures}} (\bibinfo {year} {2024}),\ \Eprint {https://arxiv.org/abs/2412.17905} {arXiv:2412.17905 [cond-mat.stat-mech]} \BibitemShut {NoStop}%
\bibitem [{Note1()}]{Note1}%
  \BibitemOpen
  \bibinfo {note} {Note that out-of-plane field does change sign under $C_{2\parallel }$ and breaks the symmetry between $\pm \phi _0$ explicitly \cite {volkov_diode} (see also Sec. \ref {subsec:memory}), that should smear the transition.}\BibitemShut {Stop}%
\bibitem [{\citenamefont {Barone}\ and\ \citenamefont {Paterno}(1982)}]{barone1982}%
  \BibitemOpen
  \bibfield  {author} {\bibinfo {author} {\bibfnamefont {A.}~\bibnamefont {Barone}}\ and\ \bibinfo {author} {\bibfnamefont {G.}~\bibnamefont {Paterno}},\ }\href {https://books.google.com/books?id=FrjvAAAAMAAJ} {\emph {\bibinfo {title} {Physics and Applications of the Josephson Effect}}},\ A Wiley-interscience publication\ (\bibinfo  {publisher} {Wiley},\ \bibinfo {year} {1982})\BibitemShut {NoStop}%
\bibitem [{\citenamefont {Tinkham}(2004)}]{tinkham2004}%
  \BibitemOpen
  \bibfield  {author} {\bibinfo {author} {\bibfnamefont {M.}~\bibnamefont {Tinkham}},\ }\href {https://books.google.com/books?id=VpUk3NfwDIkC} {\emph {\bibinfo {title} {Introduction to Superconductivity}}},\ Dover Books on Physics Series\ (\bibinfo  {publisher} {Dover Publications},\ \bibinfo {year} {2004})\BibitemShut {NoStop}%
\bibitem [{\citenamefont {Golubov}\ \emph {et~al.}(2004)\citenamefont {Golubov}, \citenamefont {Kupriyanov},\ and\ \citenamefont {Il'ichev}}]{golubov2004}%
  \BibitemOpen
  \bibfield  {author} {\bibinfo {author} {\bibfnamefont {A.~A.}\ \bibnamefont {Golubov}}, \bibinfo {author} {\bibfnamefont {M.~Y.}\ \bibnamefont {Kupriyanov}},\ and\ \bibinfo {author} {\bibfnamefont {E.}~\bibnamefont {Il'ichev}},\ }\bibfield  {title} {\bibinfo {title} {The current-phase relation in josephson junctions},\ }\href {https://doi.org/10.1103/RevModPhys.76.411} {\bibfield  {journal} {\bibinfo  {journal} {Rev. Mod. Phys.}\ }\textbf {\bibinfo {volume} {76}},\ \bibinfo {pages} {411} (\bibinfo {year} {2004})}\BibitemShut {NoStop}%
\bibitem [{\citenamefont {Savel’ev}\ \emph {et~al.}(2010)\citenamefont {Savel’ev}, \citenamefont {Yampol’skii}, \citenamefont {Rakhmanov},\ and\ \citenamefont {Nori}}]{Savelev2010}%
  \BibitemOpen
  \bibfield  {author} {\bibinfo {author} {\bibfnamefont {S.}~\bibnamefont {Savel’ev}}, \bibinfo {author} {\bibfnamefont {V.~A.}\ \bibnamefont {Yampol’skii}}, \bibinfo {author} {\bibfnamefont {A.~L.}\ \bibnamefont {Rakhmanov}},\ and\ \bibinfo {author} {\bibfnamefont {F.}~\bibnamefont {Nori}},\ }\bibfield  {title} {\bibinfo {title} {Terahertz josephson plasma waves in layered superconductors: spectrum, generation, nonlinear and quantum phenomena},\ }\href {https://doi.org/10.1088/0034-4885/73/2/026501} {\bibfield  {journal} {\bibinfo  {journal} {Reports on Progress in Physics}\ }\textbf {\bibinfo {volume} {73}},\ \bibinfo {pages} {026501} (\bibinfo {year} {2010})}\BibitemShut {NoStop}%
\bibitem [{\citenamefont {Gaifullin}\ \emph {et~al.}(1999)\citenamefont {Gaifullin}, \citenamefont {Matsuda}, \citenamefont {Chikumoto}, \citenamefont {Shimoyama}, \citenamefont {Kishio},\ and\ \citenamefont {Yoshizaki}}]{plasmon_exp_1999}%
  \BibitemOpen
  \bibfield  {author} {\bibinfo {author} {\bibfnamefont {M.~B.}\ \bibnamefont {Gaifullin}}, \bibinfo {author} {\bibfnamefont {Y.}~\bibnamefont {Matsuda}}, \bibinfo {author} {\bibfnamefont {N.}~\bibnamefont {Chikumoto}}, \bibinfo {author} {\bibfnamefont {J.}~\bibnamefont {Shimoyama}}, \bibinfo {author} {\bibfnamefont {K.}~\bibnamefont {Kishio}},\ and\ \bibinfo {author} {\bibfnamefont {R.}~\bibnamefont {Yoshizaki}},\ }\bibfield  {title} {\bibinfo {title} {$\mathit{c}$-axis superfluid response and quasiparticle damping of underdoped bi:2212 and bi:2201},\ }\href {https://doi.org/10.1103/PhysRevLett.83.3928} {\bibfield  {journal} {\bibinfo  {journal} {Phys. Rev. Lett.}\ }\textbf {\bibinfo {volume} {83}},\ \bibinfo {pages} {3928} (\bibinfo {year} {1999})}\BibitemShut {NoStop}%
\bibitem [{\citenamefont {Schmidt}(1997)}]{shmidt}%
  \BibitemOpen
  \bibfield  {author} {\bibinfo {author} {\bibfnamefont {V.}~\bibnamefont {Schmidt}},\ }\href@noop {} {\emph {\bibinfo {title} {The Physics of Superconductors: Introduction to Fundamentals and Applications}}}\ (\bibinfo  {publisher} {Springer-Verlag Berlin Heidelberg},\ \bibinfo {year} {1997})\BibitemShut {NoStop}%
\bibitem [{\citenamefont {Chakrabarti}\ and\ \citenamefont {Acharyya}(1999)}]{dyntr1}%
  \BibitemOpen
  \bibfield  {author} {\bibinfo {author} {\bibfnamefont {B.~K.}\ \bibnamefont {Chakrabarti}}\ and\ \bibinfo {author} {\bibfnamefont {M.}~\bibnamefont {Acharyya}},\ }\bibfield  {title} {\bibinfo {title} {Dynamic transitions and hysteresis},\ }\href {https://doi.org/10.1103/RevModPhys.71.847} {\bibfield  {journal} {\bibinfo  {journal} {Rev. Mod. Phys.}\ }\textbf {\bibinfo {volume} {71}},\ \bibinfo {pages} {847} (\bibinfo {year} {1999})}\BibitemShut {NoStop}%
\bibitem [{\citenamefont {Kiselev}\ \emph {et~al.}(2019)\citenamefont {Kiselev}, \citenamefont {Averkin}, \citenamefont {Fistul}, \citenamefont {Koshelets},\ and\ \citenamefont {Ustinov}}]{dyntr2}%
  \BibitemOpen
  \bibfield  {author} {\bibinfo {author} {\bibfnamefont {E.~I.}\ \bibnamefont {Kiselev}}, \bibinfo {author} {\bibfnamefont {A.~S.}\ \bibnamefont {Averkin}}, \bibinfo {author} {\bibfnamefont {M.~V.}\ \bibnamefont {Fistul}}, \bibinfo {author} {\bibfnamefont {V.~P.}\ \bibnamefont {Koshelets}},\ and\ \bibinfo {author} {\bibfnamefont {A.~V.}\ \bibnamefont {Ustinov}},\ }\bibfield  {title} {\bibinfo {title} {Two-tone spectroscopy of a squid metamaterial in the nonlinear regime},\ }\href {https://doi.org/10.1103/PhysRevResearch.1.033096} {\bibfield  {journal} {\bibinfo  {journal} {Phys. Rev. Res.}\ }\textbf {\bibinfo {volume} {1}},\ \bibinfo {pages} {033096} (\bibinfo {year} {2019})}\BibitemShut {NoStop}%
\bibitem [{Note2()}]{Note2}%
  \BibitemOpen
  \bibinfo {note} {Specifically, the Josephson current $\protect \bar {I}_J$ can be bound by $\min \left [\protect \bar {I}_c (\phi -\phi _0),\protect \bar {I}_c\right ]$. Neglecting $\protect \bar {I}_c$ one gets $(\phi -\phi _0)_{max} \sim \protect \bar {I}_0/(\protect \bar {\omega }^2+\protect \bar {\omega }/\protect \sqrt {\beta _c})$, such that (taking $\protect \bar {I}_c\sim 1$ for $\theta $ around 45$^\circ $), $I_J\ll I_0$ for all $I_0$.}\BibitemShut {Stop}%
\bibitem [{\citenamefont {Yerzhakov}\ \emph {et~al.}(2024)\citenamefont {Yerzhakov}, \citenamefont {Yeh},\ and\ \citenamefont {Balatsky}}]{yerzhakov2024}%
  \BibitemOpen
  \bibfield  {author} {\bibinfo {author} {\bibfnamefont {H.}~\bibnamefont {Yerzhakov}}, \bibinfo {author} {\bibfnamefont {T.-T.}\ \bibnamefont {Yeh}},\ and\ \bibinfo {author} {\bibfnamefont {A.}~\bibnamefont {Balatsky}},\ }\bibfield  {title} {\bibinfo {title} {Induction of orbital currents and kapitza stabilization in superconducting circuits with laguerre-gaussian microwave beams},\ }\href {https://doi.org/10.1103/PhysRevB.110.144519} {\bibfield  {journal} {\bibinfo  {journal} {Phys. Rev. B}\ }\textbf {\bibinfo {volume} {110}},\ \bibinfo {pages} {144519} (\bibinfo {year} {2024})}\BibitemShut {NoStop}%
\bibitem [{\citenamefont {Kleiner}\ and\ \citenamefont {M\"uller}(1994)}]{kleiner1994}%
  \BibitemOpen
  \bibfield  {author} {\bibinfo {author} {\bibfnamefont {R.}~\bibnamefont {Kleiner}}\ and\ \bibinfo {author} {\bibfnamefont {P.}~\bibnamefont {M\"uller}},\ }\bibfield  {title} {\bibinfo {title} {Intrinsic josephson effects in high-${\mathit{t}}_{\mathit{c}}$ superconductors},\ }\href {https://doi.org/10.1103/PhysRevB.49.1327} {\bibfield  {journal} {\bibinfo  {journal} {Phys. Rev. B}\ }\textbf {\bibinfo {volume} {49}},\ \bibinfo {pages} {1327} (\bibinfo {year} {1994})}\BibitemShut {NoStop}%
\bibitem [{\citenamefont {Kleiner}\ \emph {et~al.}(2000)\citenamefont {Kleiner}, \citenamefont {Gaber},\ and\ \citenamefont {Hechtfischer}}]{kleiner2000}%
  \BibitemOpen
  \bibfield  {author} {\bibinfo {author} {\bibfnamefont {R.}~\bibnamefont {Kleiner}}, \bibinfo {author} {\bibfnamefont {T.}~\bibnamefont {Gaber}},\ and\ \bibinfo {author} {\bibfnamefont {G.}~\bibnamefont {Hechtfischer}},\ }\bibfield  {title} {\bibinfo {title} {Stacked long josephson junctions in zero magnetic field: A numerical study of coupled one-dimensional sine-gordon equations},\ }\href {https://doi.org/10.1103/PhysRevB.62.4086} {\bibfield  {journal} {\bibinfo  {journal} {Phys. Rev. B}\ }\textbf {\bibinfo {volume} {62}},\ \bibinfo {pages} {4086} (\bibinfo {year} {2000})}\BibitemShut {NoStop}%
\bibitem [{\citenamefont {Singley}\ \emph {et~al.}(2004)\citenamefont {Singley}, \citenamefont {Abo-Bakr}, \citenamefont {Basov}, \citenamefont {Feikes}, \citenamefont {Guptasarma}, \citenamefont {Holldack}, \citenamefont {H\"ubers}, \citenamefont {Kuske}, \citenamefont {Martin}, \citenamefont {Peatman}, \citenamefont {Schade},\ and\ \citenamefont {W\"ustefeld}}]{plasma2}%
  \BibitemOpen
  \bibfield  {author} {\bibinfo {author} {\bibfnamefont {E.~J.}\ \bibnamefont {Singley}}, \bibinfo {author} {\bibfnamefont {M.}~\bibnamefont {Abo-Bakr}}, \bibinfo {author} {\bibfnamefont {D.~N.}\ \bibnamefont {Basov}}, \bibinfo {author} {\bibfnamefont {J.}~\bibnamefont {Feikes}}, \bibinfo {author} {\bibfnamefont {P.}~\bibnamefont {Guptasarma}}, \bibinfo {author} {\bibfnamefont {K.}~\bibnamefont {Holldack}}, \bibinfo {author} {\bibfnamefont {H.~W.}\ \bibnamefont {H\"ubers}}, \bibinfo {author} {\bibfnamefont {P.}~\bibnamefont {Kuske}}, \bibinfo {author} {\bibfnamefont {M.~C.}\ \bibnamefont {Martin}}, \bibinfo {author} {\bibfnamefont {W.~B.}\ \bibnamefont {Peatman}}, \bibinfo {author} {\bibfnamefont {U.}~\bibnamefont {Schade}},\ and\ \bibinfo {author} {\bibfnamefont {G.}~\bibnamefont {W\"ustefeld}},\ }\bibfield  {title} {\bibinfo {title} {Measuring the josephson plasma resonance in ${\mathrm{bi}}_{2}{\mathrm{sr}}_{2}{\mathrm{cacu}}_{2}{\mathrm{o}}_{8}$ using intense coherent thz synchrotron radiation},\ }\href
  {https://doi.org/10.1103/PhysRevB.69.092512} {\bibfield  {journal} {\bibinfo  {journal} {Phys. Rev. B}\ }\textbf {\bibinfo {volume} {69}},\ \bibinfo {pages} {092512} (\bibinfo {year} {2004})}\BibitemShut {NoStop}%
\end{thebibliography}%

\end{document}